\documentclass[preprint,preprintnumbers,amsmath,amssymb,floatfix,nofootinbib]{revtex4}
\usepackage{epsf, array, color}
\usepackage{graphicx}
\usepackage{subfigure}
\usepackage{bbold}
\usepackage{amsmath}
\usepackage{mathtools}
\usepackage{color}
\usepackage{dcolumn}

\def\pb{\rm pb}

\newcommand{\bea}{\begin{eqnarray}}
\newcommand{\eea}{\end{eqnarray}}

\def\beq{\begin{equation}}
\def\eeq{\end{equation}}

\def\gev{\rm GeV}
\def\tev{\rm TeV}

\begin{document}

\title{Singlet Model Interference Effects with High Scale UV Physics}

\author{S.~Dawson$^{\, a}$ and I.~M.~Lewis$^{\, b}$ }
\affiliation{
\vspace*{.5cm}
  \mbox{$^a$Department of Physics,\\
  Brookhaven National Laboratory, Upton, N.Y., 11973  USA}\\
 \mbox{$^b$ Department of Physics and Astronomy,\\
University of Kansas, Lawrence, Kansas, 66045 USA}
 \vspace*{1cm}}

\date{\today}

\begin{abstract}
 One of the simplest extensions of the Standard Model (SM)   is the addition of  a scalar gauge
singlet, $S$.
 If $S$ is not forbidden by a symmetry from mixing with the Standard
Model Higgs boson, the mixing will generate non-SM rates for  Higgs production
and decays.   In general, there could also be  unknown high energy physics that generates additional effective
low energy interactions. We show that interference effects between the scalar resonance of the singlet model
and the effective field theory (EFT) operators can have significant effects in the Higgs sector.
 We examine a non-$Z_2$
symmetric scalar singlet model and demonstrate that a fit to the $125~GeV$ Higgs  boson couplings 
and to limits on high mass resonances, $S$,
exhibit an interesting structure and possible large cancellations of effects between the resonance contribution and the new EFT
 interactions, that invalidate conclusions based on the renormalizable singlet model alone. 
\end{abstract}

\maketitle

\section{Introduction}
 Among the simplest  extensions of the Standard Model (SM) is the addition of a 
 gauge
singlet scalar particle, $S$.
The singlet particle  couples to SM particles through its mixing with the SM-like $125~GeV$ Higgs boson. 
In general, there can be additional
interactions  between the $S$ and the
gauge bosons, which can be parameterized as effective field theory (EFT)  dimension-$5$ couplings.  
The source of these effective interactions is not relevant for our discussion and our focus is on the consequences of
the interference effects between the heavy scalar resonance and the EFT operators.  Since there are a relatively few
number of EFT operators coupling the singlet to the $SU(3)\times SU(2)\times U(1)$ gauge bosons, it is possible
to obtain interesting limits on the theory, despite the addition of new parameters.

In the absence of a $Z_2$ symmetry, the singlet model allows  cubic and linear self-coupling terms   in the 
scalar potential  and a strong first order electroweak  phase transition is possible for certain values
of the parameter space\cite{Espinosa:2011ax,Profumo:2007wc,Profumo:2014opa,Curtin:2014jma,Perelstein:2016cxy},
making this theory highly motivated phenomenologically. 
 We begin by examining restrictions on the parameters of the non-$Z_2$ symmetric model from the measured $125~GeV$
Higgs couplings and from the requirement
that the electroweak minimum be the absolute minima of the potential.  We then include LHC limits on heavy resonances that
decay into SM particles (assuming that there are no additional light particles).   Novel
features of our analysis are the insistence that the parameters satisfy the minimization condition 
of the potential and our inclusion of interference
effects between the SM contributions to the Higgs widths and the contributions from the EFT interactions. 
These interference effects can be large and significantly change the allowed regions of parameter space. 

In Sec. \ref{sec:model}, we review the singlet model and the EFT interactions, along with compact expressions
for the decay widths.  Sec. \ref{sec:h1lims} discusses constraints from the $125$ GeV Higgs, and Sec. 
\ref{sec:results} contains our limits on the properties of both the $125$  scalar and EFT coefficients,
and a discussion of the size of the allowed mixing between the SM-like and heavy scalars in the presence of EFT coefficients.
Section V contains some conclusions. 
\section{Model Considerations}
\label{sec:model}
\subsection{Singlet plus EFT Model}
\label{sec:Model}
We consider a model containing the SM Higgs doublet, $H$, and an additional
scalar singlet,  $S$.
The most general renormalizable scalar potential is,
\begin{eqnarray}
V(H,S) & =& 
- \mu^2 \, H^\dagger H + \lambda (H^\dagger H)^2 +\frac{a_1}{2} \, H^\dagger H \, S
   + \frac{a_2}{2} \, H^\dagger H \, S^2\
   \nonumber \\
&&   +
b_1 S + \frac{b_2}{2} S^2 + \frac{b_3}{3} S^3 + \frac{b_4}{4} S^4.
\label{potential}
\end{eqnarray}
The singlet model has been examined in some detail
in the literature\cite{O'Connell:2006wi,Barger:2008jx,Profumo:2007wc,Pruna:2013bma,Chen:2014ask,Espinosa:2011ax,Robens:2016xkb,Dawson:2015haa,Costa:2015llh} and so our discussion is appropriately brief. 
If there is a $Z_2$ symmetry  $S\rightarrow -S$, then $a_1=b_1=b_3=0$.  The $Z_2$ non-symmetric model is, however, particularly
interesting since it is possible to arrange the parameters in such a way as to obtain a strong first
order phase transition\cite{Profumo:2007wc,Profumo:2014opa,Ahriche:2007jp,Espinosa:2011ax,Curtin:2014jma,Perelstein:2016cxy}.

 The neutral scalar components of the doublet $H$  and singlet $S$ are  denoted by $\phi_0=(h+v)/\sqrt{2}$
 and $S=s+ x $, where the vacuum expectation values are $\langle \phi_0\rangle = {v\over \sqrt{2}}$
 and $\langle S\rangle =x$. 
We require that the  global minimum of the potential correspond to the electroweak symmetry breaking (EWSB) minimum, $v=v_{EW}=246$~GeV\cite{Chen:2014ask,Espinosa:2011ax}, which places significant constraints
on the allowed parameters. 
Note that a shift of the singlet field by $S\rightarrow S+\Delta_S$ is just
a redefinition of the parameters of Eq. \ref{potential} and we are free to choose our electroweak
symmmetry breaking minimum as $(v,x)\equiv(v_{EW},0)$\footnote{This freedom to set $x=0$ does not occur
in the $Z_2$ symmetric case.}.

The physical scalars are mixtures of $h$ and $s$, and the scalar mixing is parameterized as,
\begin{eqnarray}
\begin{pmatrix}h_1 \\ h_2\end{pmatrix}=\begin{pmatrix} \cos\theta&\sin\theta \\ -\sin\theta & \cos\theta\end{pmatrix} \begin{pmatrix}h\\s\end{pmatrix},
\label{singmod}
\end{eqnarray}
where $h_{1,2}$ are the mass eigenstates with masses $m_{1}, m_2$.  
The parameters of the scalar potential can be solved for in terms of the physical masses and mixing,
\begin{eqnarray}
a_1&=&\frac{m_1^2-m_2^2}{v}\sin2\theta,\nonumber\\
b_2+\frac{a_2}{2}v^2&=&m_1^2\sin^2\theta+m_2^2\cos^2\theta,\nonumber\\
\lambda&=&\frac{m_1^2\cos^2\theta+m^2_2\sin^2\theta}{2v^2}\nonumber\\
\mu^2&=&\lambda v^2\nonumber\\
b_1&=&-\frac{v^2}{4}a_1.
\label{paramdefs}
\end{eqnarray}
Our free parameters are then,
\begin{eqnarray}
m_1=125~{\rm GeV},\;m_2,\;\theta,\;v_{EW}=246~{\rm GeV},\;x=0,\; a_2,\;b_3,\;b_4.
\end{eqnarray}
The couplings of the $h_1$ to SM particles are suppressed by $\cos\theta$ and both ATLAS
and CMS have obtained limits from the measured couplings.  ATLAS finds at $95\%$ confidence level,
$\sin\theta \le .35$, assuming no branching ratio to invisible particles\cite{Aad:2015pla}.   Using the
fitted global signal strength for the SM Higgs boson, $\mu=1.03^{+0.17}_{-0.15} $\cite{ATLAS-CONF-2015-044}, a $95\%$ confidence level
limit can be extracted, $\sin\theta \le .51$.   In the absence of the EFT coefficients, a fit to the oblique parameters also
restricts $\sin\theta$\cite{Profumo:2007wc,Dawson:2009yx,Pruna:2013bma,Chen:2014ask},
 but the limit from Higgs coupling measurements is stronger. 

The limits  on $\sin\theta$ can be significantly altered, however,  when the EFT operators are included.
 We postulate the $SU(3)\times SU(2)\times U(1)$
gauge invariant effective interactions,
\begin{equation}
L=g_s^2 {c_{gg}\over \Lambda} S G^{\mu\nu,A}G^A_{\mu\nu}
+{c_{WW}\over \Lambda} g^2 S W^{\mu\nu,a}W^a_{\mu\nu}
+{c_{BB}\over \Lambda} {g'}^2 S B^{\mu\nu}B_{\mu\nu}\, ,
\label{eff}
\end{equation}
that are assumed to arise from unknown UV physics at a scale $\Lambda$.
The scalar couplings 
to gauge bosons are suppressed by the appropriate factor of $\cos\theta$ or $\sin\theta$ and
receive  additional contributions from the interactions of Eq. \ref{eff}.   There is an interplay of effects
between the singlet-SM mixing of Eq. \ref{singmod} and the EFT contributions from Eq.~\ref{eff}, which 
requires that we fit the data to the complete  model\cite{Bauer:2016hcu,Cheung:2016trn}.

\begin{figure}
\begin{centering}
\includegraphics[width=0.3\textwidth,clip]{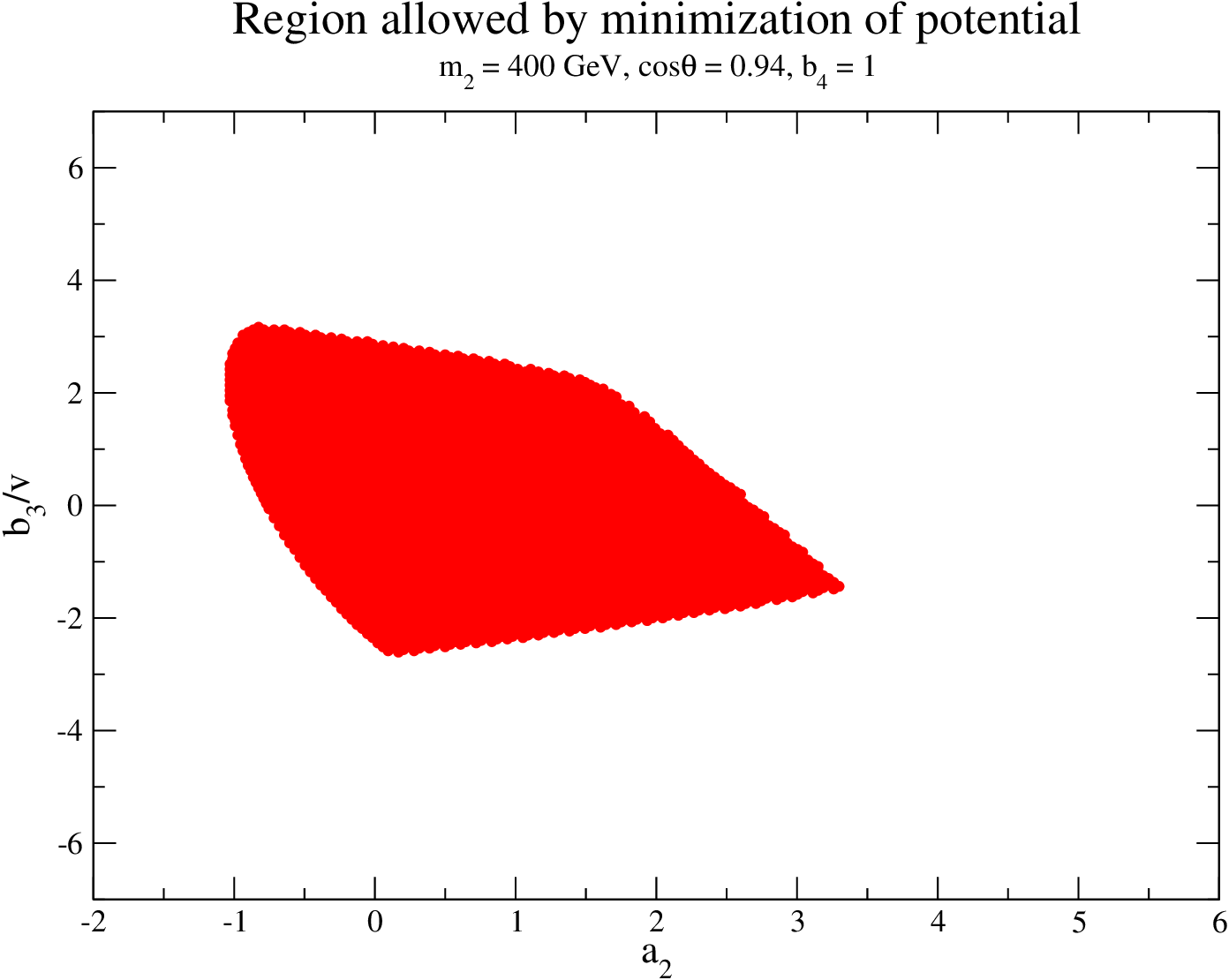}
\includegraphics[width=0.3\textwidth,clip]{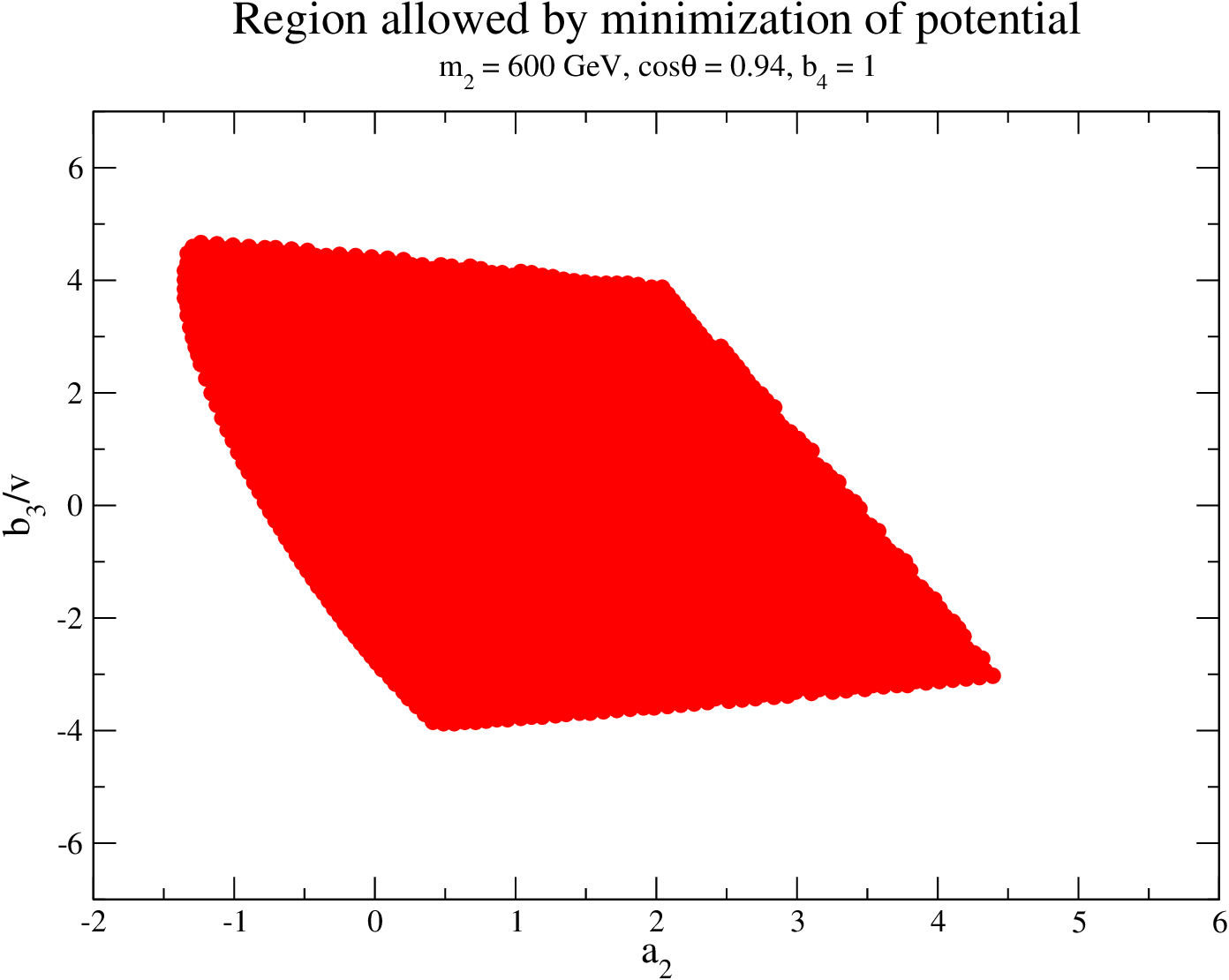}
\includegraphics[width=0.3\textwidth,clip]{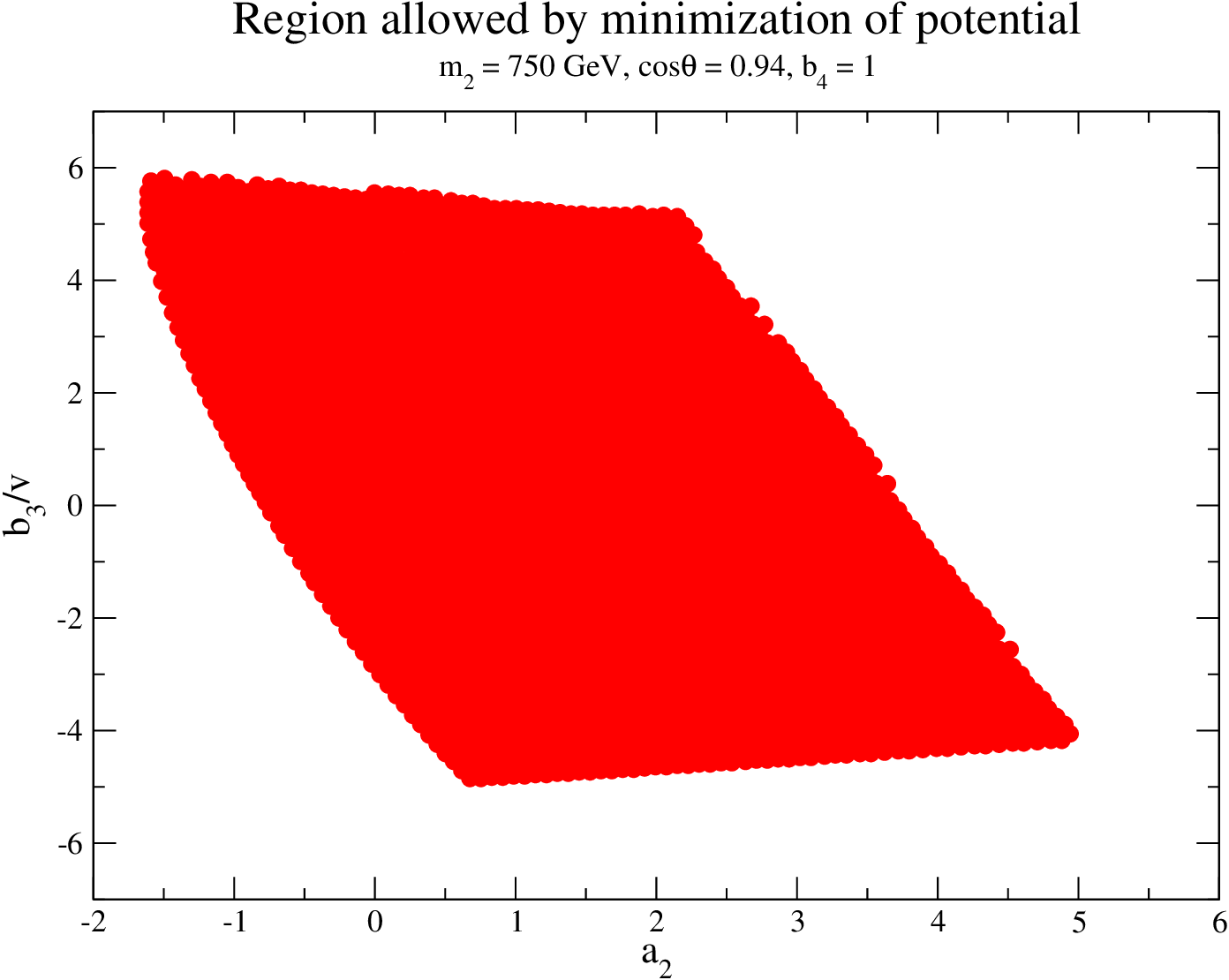}
\par\end{centering}
\caption{Regions allowed by the requirement that the electroweak minimum be
a global  minimum for $\cos \theta=0.94$, $b_4=1$ and $m_2=400,~600$, and $750~\gev$\cite{Chen:2014ask}.}
\label{fig:pot1}
\end{figure}

Finally, we need the self-interactions of the Higgs bosons
in the basis of the mass eigenstates $h_1$ and $h_2$,
\bea
V_{\rm self}&\supset& {\lambda_{111} \over 3!} h_1^3 + {\lambda_{211} \over 2!} h_2 h_1^2 
 + ...
\eea
where\cite{Pruna:2013bma,Chen:2014ask},
\bea
\lambda_{111}&=& 2 s_\theta^3 b_3+{3 a_1 \over 2}s_\theta c_\theta^2+3 a_2 s_\theta^2 c_\theta
 v + 6 c_\theta^3\ \lambda v,\nonumber \\
\lambda_{211}&=& 2 s_\theta ^2c_\theta b_3+{a_1 \over 2}c_\theta (c_\theta^2-2s_\theta^2)
+(2c_\theta ^2-s_\theta ^2)s_\theta v a_2 -6\lambda s_\theta c_\theta ^2v
\label{self}
\, .
\end{eqnarray}
and we abbreviate $s_\theta=\sin\theta$, $c_\theta=\cos\theta$ and assume $\sin\theta>0$.   In the small angle limit, to ${\cal O}(s_\theta^2)$,
\bea
\lambda_{111}&\rightarrow & 6\lambda v+{3\over 2} a_1s_\theta+ 3v s_\theta ^2(a_2-3\lambda)\\
&\sim &  {3m_1^2\over v} +s_\theta^2{3\over 2 v} \biggl(
m_2^2-4m_1^2+2a_2v^2\biggr)
\nonumber \\
\lambda_{211}&\rightarrow & {a_1\over 2} +s_\theta  v(-6\lambda +2a_2)+{s_\theta ^2\over 4}(8b_3-7a_1)\nonumber \\ &\sim &
s_\theta\biggl(-{3m_1^2\over v}+2 v a_2\biggr)
+{s_\theta c_\theta\over 2 v }(m_1^2-m_2^2)
+2b_3s_\theta^2
\, .
\eea
The restrictions on the parameters of the potential due to the requirement that the electroweak minimum be
a global  minimum were examined in Ref. \cite{Chen:2014ask,Espinosa:2011ax}.  
In Fig. \ref{fig:pot1}, we fix $b_4=1$, $\cos\theta=.94$  and show the allowed regions for $a_2$ and $b_3$
for different values of the heavy scalar mass, $m_2$.  The areas of these regions increase with $b_4$
 and the edges of the contours are completely fixed by the global
minimum requirement as described in Ref. \cite{Chen:2014ask}\footnote{${\it{i.e.}}$ all points within the shaded regions are allowed
by the minimization of the potential.}.  The regions become somewhat larger as $m_2$ increases for fixed $b_4$.
In the softly broken $Z_2$ scenario of Ref. \cite{Perelstein:2016cxy}, a first order electroweak phase transition requires
$a_2> \sim 9$.  In the model without
a $Z_2$ symmetry, a strong first order electroweak phase transition appears to be possible for 
$a_2\sim 1-2$, and negative $b_3$\cite{Profumo:2014opa}, although the maximum $m_2$ studied in this reference is $250~GeV$.  

The partial width of $h_2\rightarrow h_1h_1$ is,
\begin{eqnarray}
\Gamma(h_2\rightarrow h_1 h_1)=\frac{\lambda^2_{211}}{32\pi m_2}\sqrt{1-\frac{4 m_1^2}{m_2^2}}.
\end{eqnarray}
\begin{figure}
\begin{centering}
\includegraphics[width=0.3\textwidth,clip]{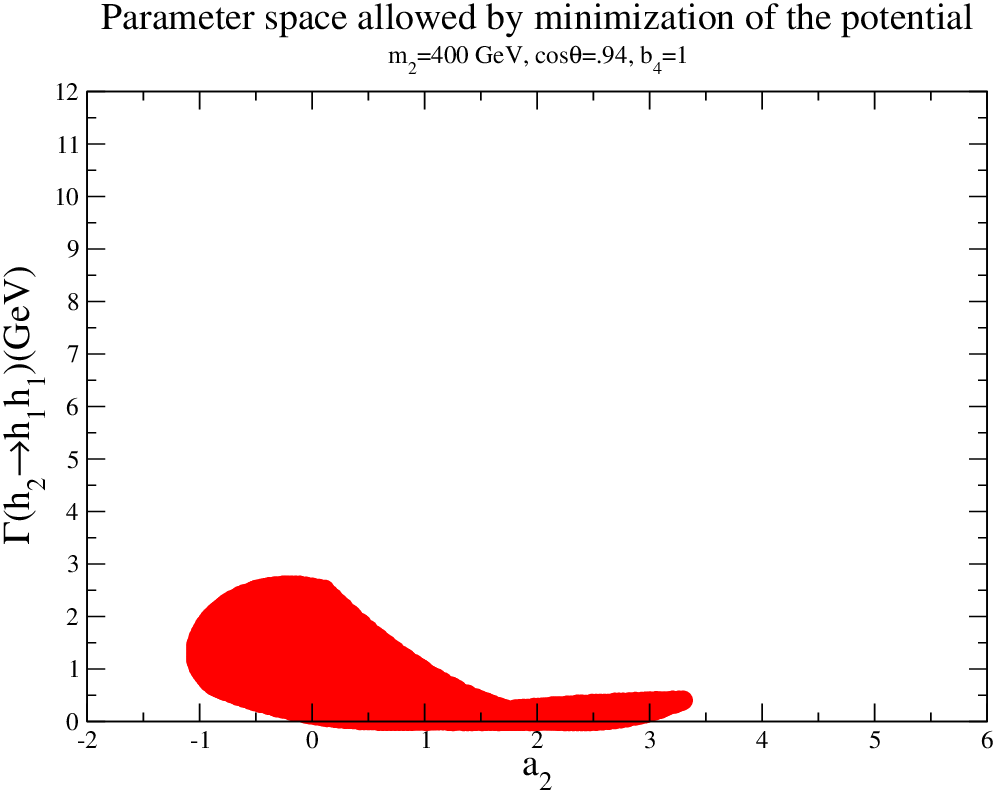}
\includegraphics[width=0.3\textwidth,clip]{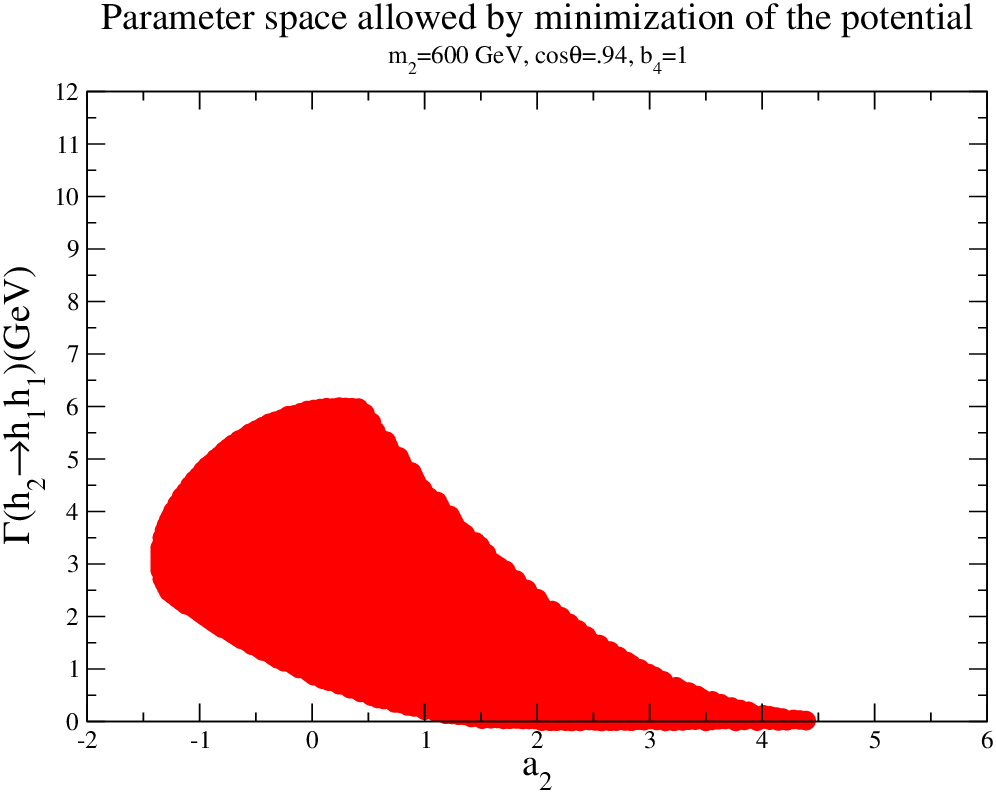}
\includegraphics[width=0.3\textwidth,clip]{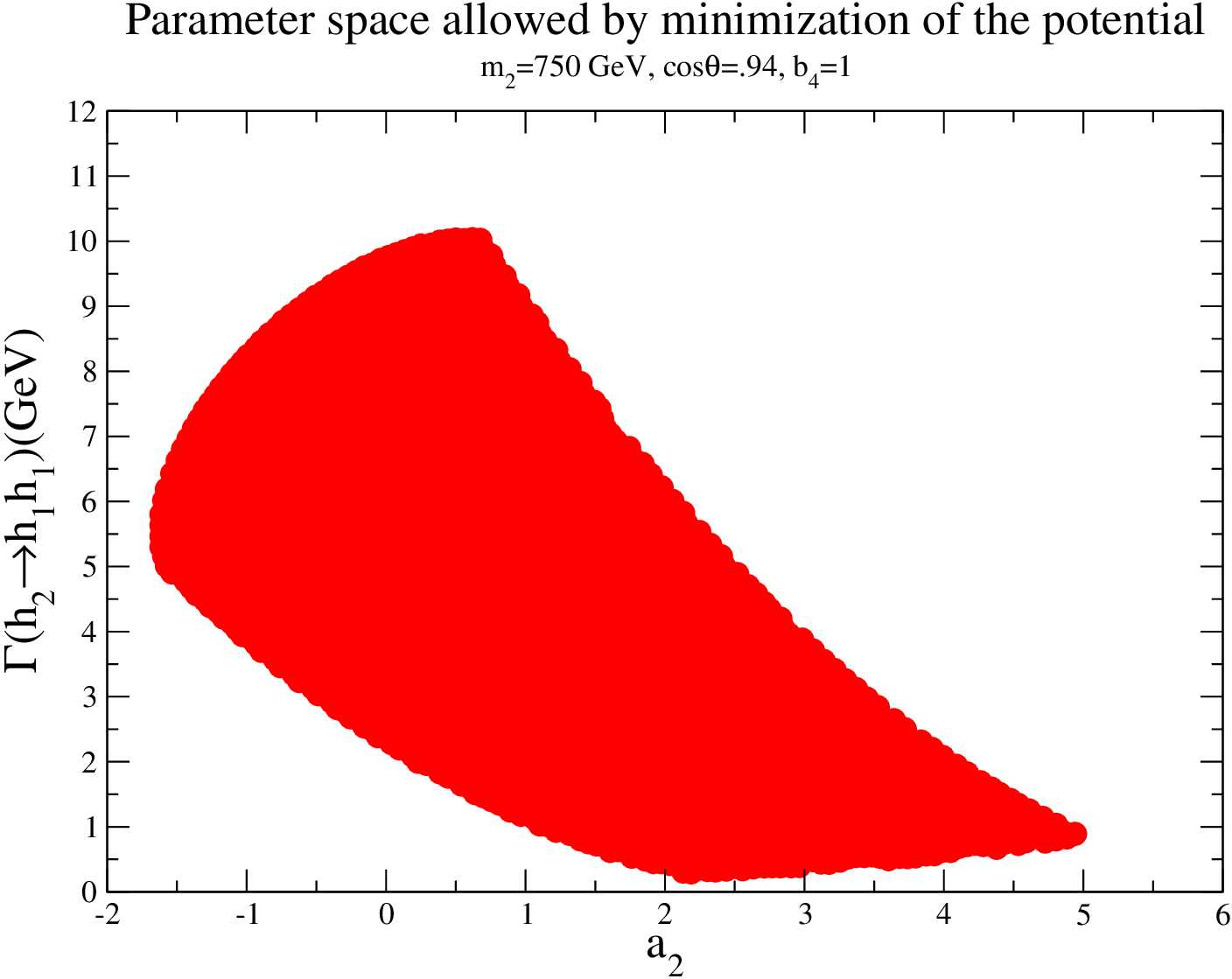}
\par\end{centering}
\caption{Allowed decay widths for $h_2\rightarrow h_1 h_1$ assuming the
parameters correspond to a global minimum of the potential for $b_4=1$, $\cos\theta=0.94$, and $m_2=400,~600$ and $750~GeV$.}
\label{fig:wid211}
\end{figure}
In Fig. \ref{fig:wid211} we show the partial widths for $h_2\rightarrow h_1 h_1$ using
the allowed values of  $b_3$ from Fig. \ref{fig:pot1} for each parameter point for representative values of the parameters.  
The width can potentially increase significantly as the resonance mass increases.  A measurement of the coupling $\lambda_{211}$ to sufficient precision could shed light on the values of $a_2$ and $b_3$.  We note that in all cases, $\Gamma(h_2\rightarrow h_1h_1)_{max}/m_2
\sim 1\%$, and so we are in a narrow width scenario.

\subsection{Results for Decay widths} 
The decays of $h_1$ and $h_2$ are affected by the SM doublet-singlet mixing and by the EFT operators. 
 Retaining the interference with the SM contributions,
we find for the heavier state: 
\begin{eqnarray}
\Gamma(h_2\rightarrow \gamma \gamma)&=&{e^4 m_2^3\over 4\pi}
\mid\sin\theta \biggl({\Sigma_i N_{ci}e_i^2F_i(\tau_{2i})\over 32\pi^2 v}\biggr)
+\cos\theta{ c_{\gamma \gamma}\over\Lambda}\mid^2\nonumber\\
\Gamma(h_2\rightarrow gg)&=&{2 g_s^4 m_2^3\over \pi}
\mid \sin\theta {\Sigma_i F_i(\tau_{2i})\over 64\pi^2 v}
+\cos\theta {c_{gg}\over\Lambda}\mid^2\nonumber\\
\Gamma(h_2\rightarrow ZZ)&=&\frac{1}{32\pi}\frac{m_2^3}{v^2}\sqrt{1-4 x_{2Z}}
\left\{2^7\cos^2\theta \frac{c^2_{ZZ} M_Z^4}{\Lambda^2v^2}\left(1-4 x_{2Z}+6 x_{2Z}^2\right)\right.\nonumber\\
&&\left.+3\cdot 2^5 \cos\theta\sin\theta \frac{c_{ZZ} M^2_Z}{v\Lambda}x_{2Z}(1-2x_{2Z})
+\sin^2\theta\left(1-4 x_{2Z}+12 x^2_{2Z}\right)\right\}\nonumber\nonumber\\
\Gamma(h_2\rightarrow Z\gamma)&=& {e^4 m_2^3
\over  2\pi s_W^2c_W^2} (1-x_{2Z})^3\mid  \sin\theta {c_W s_W\over 32\pi^2 v}(A_F+A_W)-\cos\theta {c_{z\gamma}\over\Lambda}\mid^2\nonumber \\
\Gamma(h_2\rightarrow W^+W^-)&=&\frac{1}{16}\frac{m_2^3}{\pi v^2}\sqrt{1-4 x_{2W}}
\left\{2^7\cos^2\theta \frac{c^2_{WW} M_W^4}{\Lambda^2 v^2}\left(1-4 x_{2W}+6 x_{2W}^2\right)\right.\nonumber\\
&&\left.+3\cdot 2^5 \cos\theta\sin\theta \frac{c_{WW} M^2_W}{v\Lambda}x_{2W}(1-2x_{2W})+\sin^2\theta\left(1-4 x_{2W}+12 x^2_{2W}\right)\right\}\nonumber\\
\Gamma(h_2\rightarrow f {\overline f})&=& \sin^2\theta \Gamma(h\rightarrow f {\overline f})_{SM}\, ,
\label{h2dec}
\end{eqnarray}
where~\cite{Gunion:1989we,Djouadi:2005gi,Contino:2014aaa},
\begin{eqnarray}
F_i(\tau_{2i})&=& -2\tau_{2i}\biggl(1+(1-\tau_{2i})f(\tau_{2i})\biggr)~{\rm for~fermions}\nonumber \\
F_W(\tau_{2W})&=& 2+ 3\tau_{2W}+3\tau_{2W}(2-\tau_{2W})f(\tau_{2W})~{\rm for~gauge~bosons}\nonumber \\
x_{iV}&=&{M_V^2\over m_i^2}\nonumber \\
c_{\gamma\gamma}&=& c_{WW}+c_{BB}\nonumber\\
c_{ZZ}&=&  c_W^4c_{WW}+s_W^4c_{BB}\nonumber\\
c_{Z\gamma}&=&c_{BB} s^2_W-c_{WW}c^2_W\, ,
\label{defs}
\end{eqnarray}
and $e_i$ is the electric charge of particle $i$,  $c_W=M_W/M_Z$, $N_{ci}=3(1)$ for quarks (leptons), 
$\tau_{2i}={4M_{i}^2\over m_2^2}$, $M_{i}$ is the mass of the
appropriate fermion or the $W$ boson,  $A_F$ and $A_W$ are given in Ref.\cite{Gunion:1989we}, 
and
\begin{eqnarray}
f(\tau)=&\biggl[\sin^{-1}\biggl({1\over\sqrt{\tau}}\biggr)\biggr]^2, & {\hbox{if~}}\tau \ge 1\nonumber \\
=& -{1\over 4} \biggl[\ln\biggl(
{1+\sqrt{1-\tau}\over 1-\sqrt{1-\tau}}\biggr)-i\pi\biggr]^2
&{\hbox{if~}}\tau <1\, .
\end{eqnarray}

If we consider a model with no mixing with the SM Higgs,  $\sin\theta=0$, we have
approximately, 
\begin{eqnarray}
\Gamma(h_2\rightarrow \gamma\gamma)&=&.04 c_{\gamma\gamma}^2
\biggl({m_2\over 600~GeV}\biggr)^3
\biggl({2~TeV\over \Lambda(TeV)}\biggr)^2~\gev\nonumber \\
\Gamma(h_2\rightarrow W^+W^-)&=&0.15c_{WW}^2
\biggl({m_2\over 600~GeV}\biggr)^3
\biggl({2~TeV\over \Lambda(TeV)}\biggr)^2~\gev\nonumber \\
\Gamma(h_2\rightarrow ZZ)&=&1.2 c_{ZZ}^2
\biggl({m_2\over 600~GeV}\biggr)^3
\biggl({2~TeV\over \Lambda(TeV)}\biggr)^2~\gev\nonumber \\
\Gamma(h_2\rightarrow Z\gamma)&=&0.43 c_{Z\gamma}^2
\biggl({m_2\over 600~GeV}\biggr)^3
\biggl({2~TeV\over \Lambda(TeV)}\biggr)^2~\gev\, .
\label{clim}
\end{eqnarray}
Note that Eq. \ref{clim}  is an over-constrained result due to the relations of Eq. \ref{defs}.

The lighter Higgs boson  ($m_1=125~GeV$) decay widths  are,
 \begin{eqnarray}
\Gamma(h_1\rightarrow gg)&=&{2 g_s^4 m_1^3\over \pi}
\mid -\cos\theta {\Sigma_i F_i(\tau_{1i})\over 64\pi^2 v}
+\sin\theta {c_{gg}\over\Lambda}\mid^2\nonumber\\
\Gamma(h_1\rightarrow \gamma \gamma)&=&{e^4 m_1^3\over 4\pi}
\mid-\cos\theta \biggl({\Sigma_i N_{ci}e_i^2F_i(\tau_{1i})\over 32\pi^2 v}\biggr)
+\sin\theta{ c_{\gamma \gamma}\over\Lambda}\mid^2\nonumber\\
\Gamma(h_1\rightarrow WW^*)&=&\frac{18 g^2 M_W^4}{\pi^3v^2 m_1}
\left\{\sin^2\theta \frac{c^2_{WW}}{v^2\Lambda^2 }m_1^4 I_3(M_W)-\cos\theta\sin\theta \frac{c_{WW}}{4 v\Lambda }m_1^2 I_2(M_W)+\frac{1}{64}\cos^2\theta I_1(M_W)\right\}\nonumber\\
\Gamma(h_1\rightarrow ZZ^*)&=&\kappa\frac{2 g^2 M_Z^4}{c_W^2\pi^3v^2 m_1}\left\{\sin^2\theta \frac{c^2_{ZZ}}{v^2\Lambda^2
}m_1^4 I_3(M_Z)-\cos\theta\sin\theta \frac{c_{ZZ}}{4 v\Lambda}m_1^2 I_2(M_Z)+\frac{1}{64}\cos^2\theta I_1(M_Z)\right\}\nonumber\\
\Gamma(h_1\rightarrow Z\gamma)&=& {e^4 m_1^3
\over 2 \pi s_W^2c_W^2} (1-x_{1Z})^3\mid  \cos\theta {c_W s_W\over 32\pi^2 v}(A_F+A_W)+\sin\theta {c_{z\gamma}\over\Lambda}\mid^2\nonumber \\
\Gamma(h_1\rightarrow f {\overline f})&=& \cos^2\theta \Gamma(h\rightarrow f {\overline f})_{SM}
\label{h1wid}
\end{eqnarray}
where,
\begin{eqnarray}
I_1(M_W)&=&\int^{(m_1-M_W)^2}_0 dq^2 \frac{q^2}{m_1^2}\left(1+\frac{1}{3}\frac{{\hat{\lambda}}(m_1^2,M_W^2,q^2)}{4 q^2 M_W^2}\right)\frac{{\hat{\lambda}}^{1/2}(m_1^2,M_W^2,q^2)}{(q^2-M_W^2)^2+\Gamma_W^2 M_W^2}\nonumber\\
I_2(M_W)&=&\int^{(m_1-M_W)^2}_0 dq^2 \frac{q^2}{m_1^2}\frac{M_1^2-M_W^2-q^2}{2 m_1^2}\frac{{\hat{\lambda}}^{1/2}(m_1^2,M_W^2,q^2)}{(q^2-M_W^2)^2+\Gamma_W^2 M_W^2}\\
I_3(M_W)&=&\int^{(m_1-M_W)^2}_0 dq^2 \frac{q^2}{m_1^2}\frac{3(m_1^2-M_W^2-q^2)^2
-{\hat{\lambda}}(m_1^2,M_W^2,q^2)}{12 m_1^4}\frac{{\hat{\lambda}}^{1/2}(m_1^2,M_W^2,q^2)}{(q^2-M_W^2)^2+\Gamma_W^2 M_W^2}\nonumber\\
{\hat{\lambda}}(x,y,z)&=&(x-y-z)^2-4yz\nonumber\, ,
\end{eqnarray}
 $\tau_{1i}={4M_{i}^2\over m_1^2}$, the coefficient $\kappa$ is,
\begin{eqnarray}
\kappa&=&3({1\over 2}-s_W^2)^2+s_W^4)+3N_c((-{1\over 2}+{1\over 3}s_W^2)^2+{1\over 9}s_W^4)
+2N_c(({1\over 2}-{2\over 3}s_W^2)^2+{4\over 9}s_W^4)
\nonumber\\
&=&3.68,
\end{eqnarray}
with $N_c=3$ and $s_W^2=\sin^2\theta_W=1-{M_W^2\over M_Z^2}$.

\begin{figure}
\begin{centering}
\includegraphics[width=0.4\textwidth,clip]{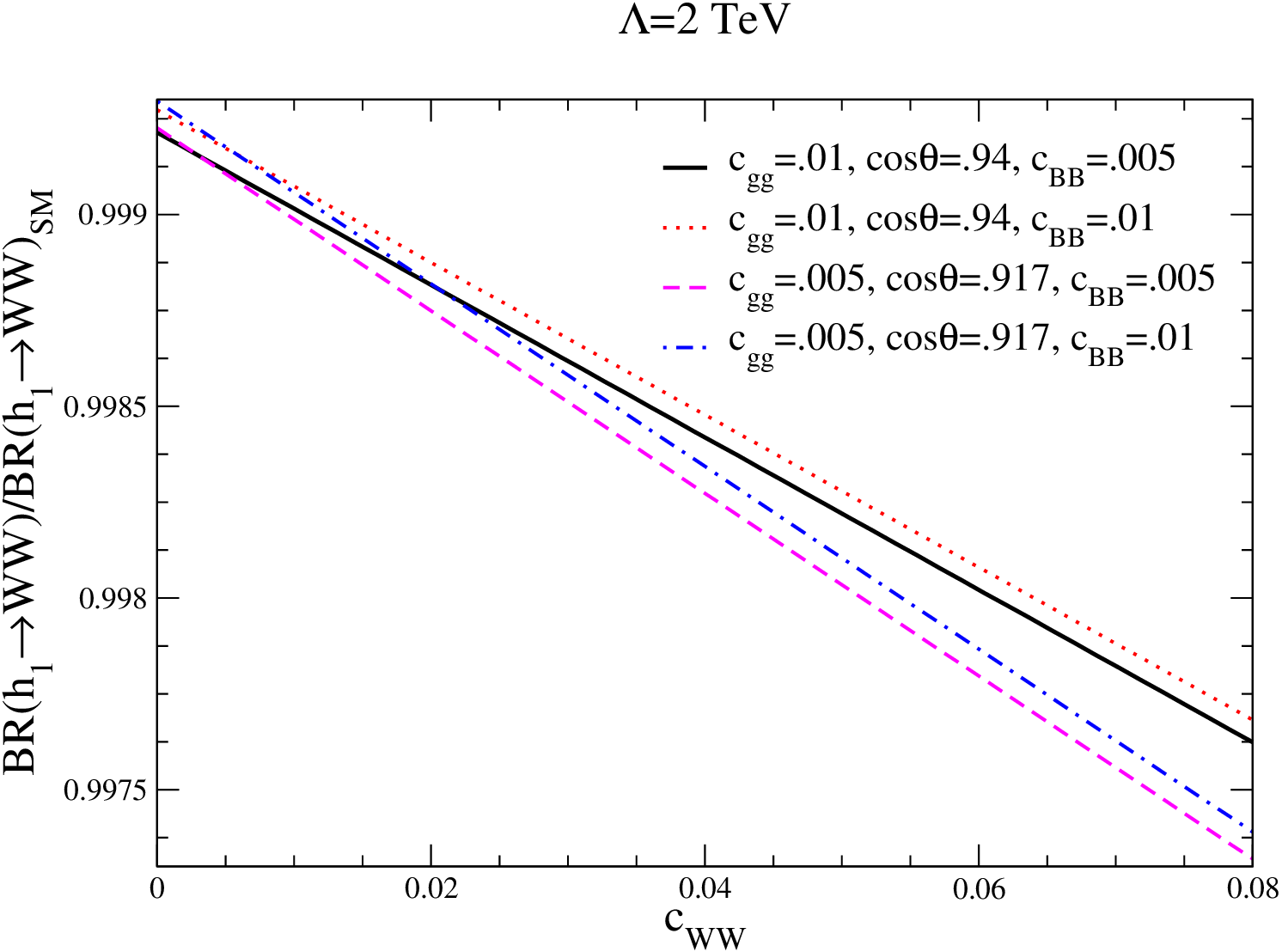}
\includegraphics[width=0.4\textwidth,clip]{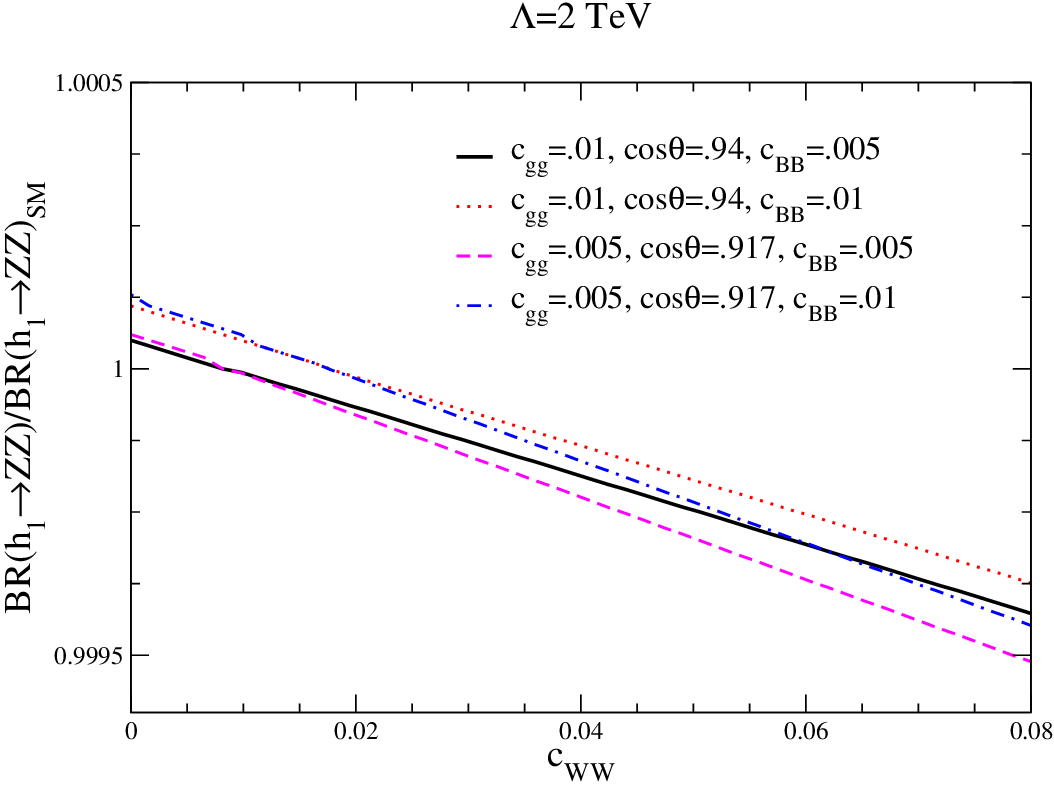}
\par\end{centering}
\caption{Branching ratio for (LHS) $h_1\rightarrow WW$, and (RHS) $h_1\rightarrow ZZ$ for representative
values of the parameters.}
\label{fig:h1ww}
\end{figure}

\begin{figure}
\begin{centering}
\includegraphics[width=0.4\textwidth,clip]{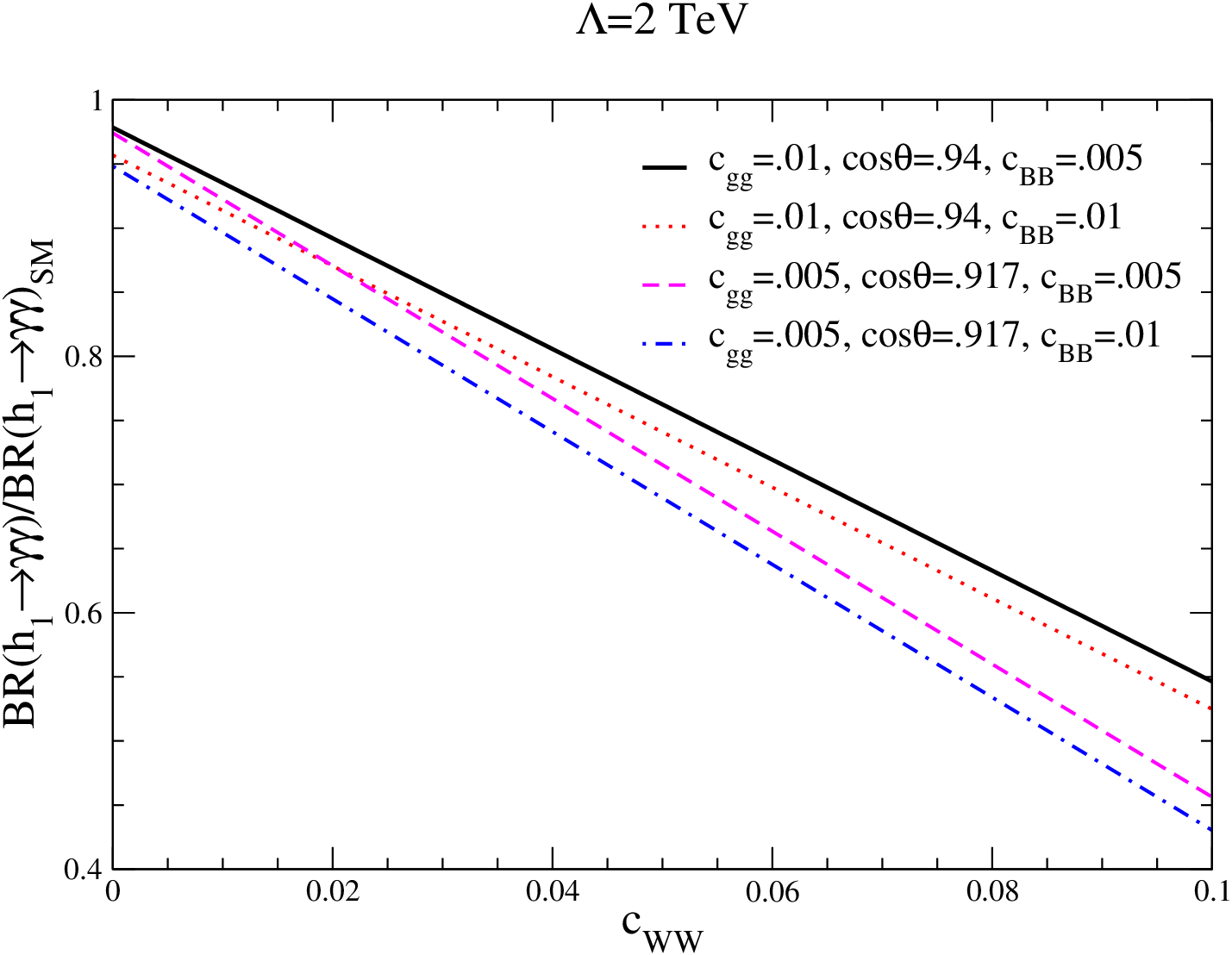}
\includegraphics[width=0.4\textwidth,clip]{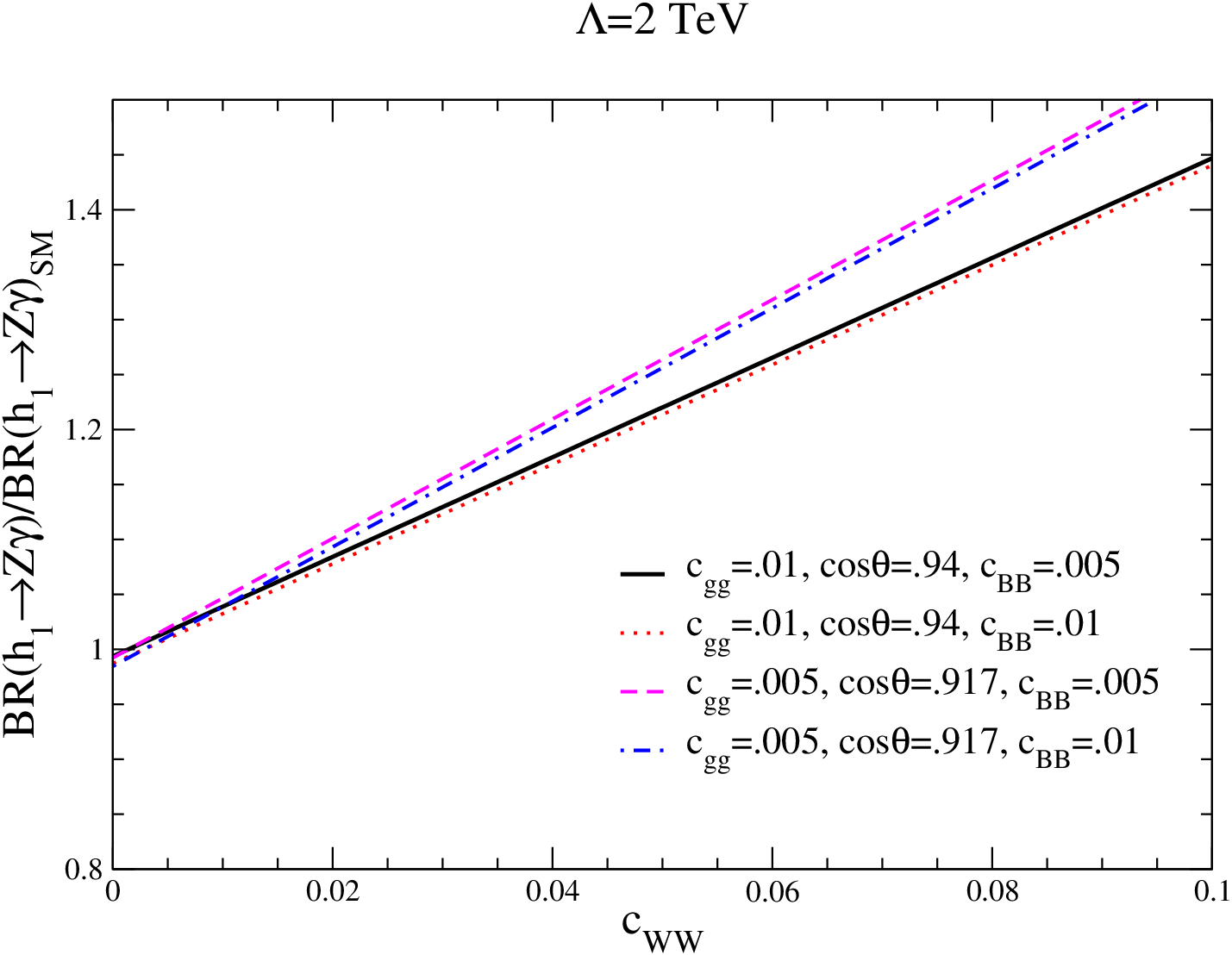}
\par\end{centering}
\caption{Branching ratios for (LHS) $h_1\rightarrow \gamma\gamma$, and (RHS) $h_1\rightarrow Z\gamma$ for 
representative values of the parameters.}
\label{fig:h1gg}
\end{figure}

Some typical branching ratios of $h_1$  into $WW$ and $ZZ$ normalized to the SM are shown in Fig. \ref{fig:h1ww}, and demonstrate little
sensitivity to either $c_{BB}$ or $c_{WW}$ with sub-percent level deviations.  
The branching ratios to $\gamma\gamma$ and $Z\gamma$ are shown in Fig. \ref{fig:h1gg}
and are  very sensitive to $c_{WW}$ and $c_{BB}$, changing upwards of 50\% from the SM values. This is due to the SM rate first occuring at one loop.
 We note that in the limit $c_{gg}=c_{WW}=c_{BB}=0$, all 
of the branching ratios are equal to their SM values for $\sin\theta=0$, and the deviations from $1$ in Figs. \ref{fig:h1ww} and \ref{fig:h1gg}
are a result of the interplay between the singlet mixing and the EFT operators. These figures retain only the
linear terms in the EFT couplings, as we have implicitly assumed $s_\theta$ is small and we note that the $c_i^2$ coefficients
are always suppressed by $s_\theta^2$ for $h_1$ production (see Eq.  \ref{h1wid}). 

For completeness, we note that the hadronic cross section  for production of $h_1$ or $h_2$ from gluon
fusion is,
\begin{equation}
\sigma(pp\rightarrow h_i)={\pi^2\over 8 m_i S_H}\Gamma(h_i\rightarrow gg)L
\end{equation}
where
\begin{equation}
L=\int_{\ln(\sqrt{\zeta})}^{-\ln(\sqrt{\zeta})}
dy g(\sqrt{\zeta}e^y)g(\sqrt{\zeta} e^{-y})\, ,
\end{equation}
$\sqrt{S_H}$ is the hadronic center-of-mass energy and $\zeta=m_i^2/S_H$.

\section{Constraints from $h_1$}
\label{sec:h1lims}

The measurements of SM Higgs couplings place stringent restrictions on the allowed 
parameters of the model.
Both ATLAS and CMS  limit the mixing angle, $\theta$, in the singlet model 
in the case $c_{gg}=c_{WW}=c_{BB}=0$, as discussed in the previous section.
These limits are significantly effected by the addition of the EFT operators. 
We  fit to the parameters of our model using the combined ATLAS/CMS $8~TeV$
results\cite{ATLAS-CONF-2015-044}.  The simplest possible 
limit is obtained by a fit to
the over-all gluon fusion signal strenth for $h_1$,
\begin{equation}
\mu_{ggF}=1.03^{+.17}_{-.15}\, .
\end{equation}

\begin{figure}
\begin{centering}
\includegraphics[width=0.5\textwidth]{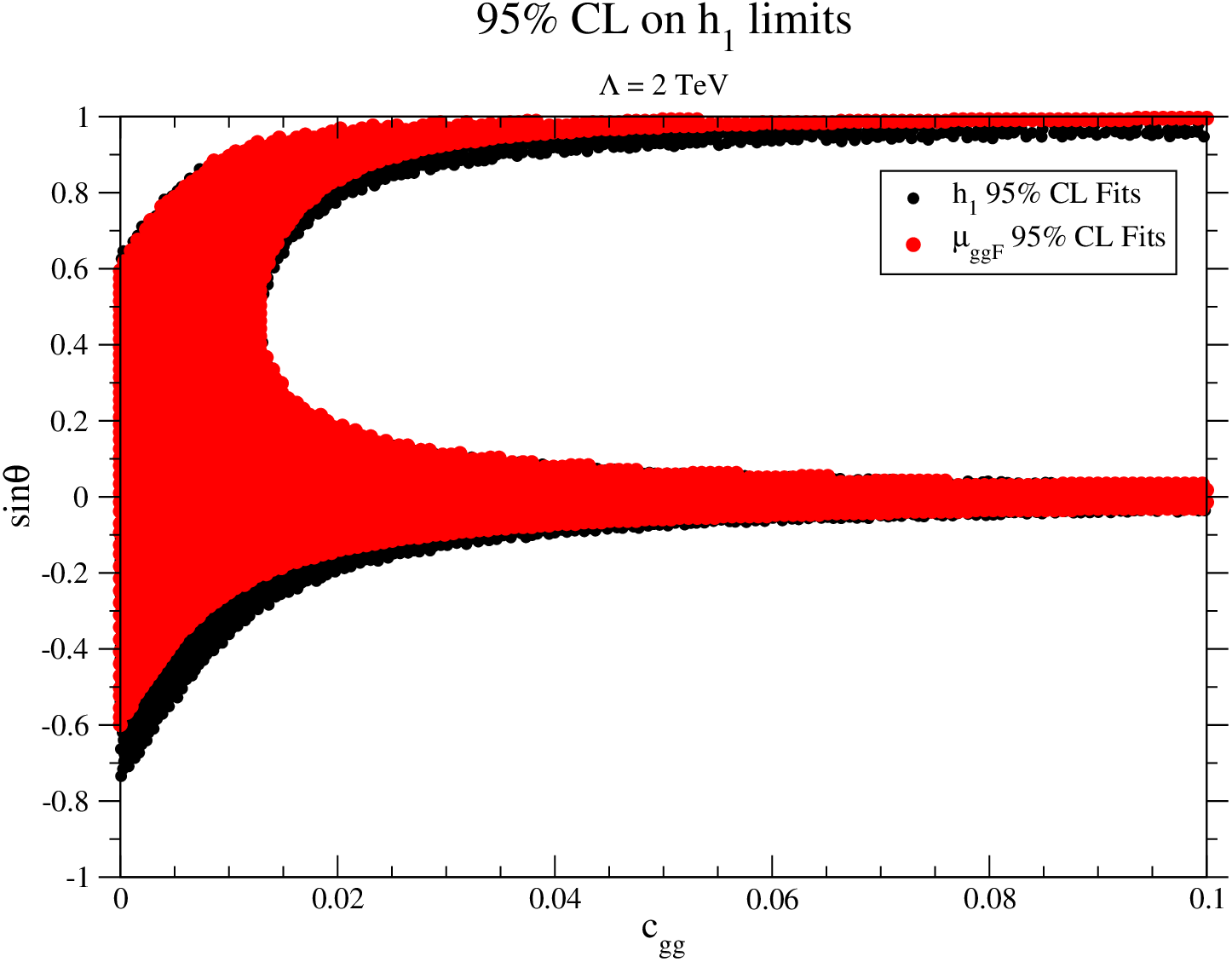}
\par\end{centering}
\caption{95\% confidence level allowed regions using the gluon fusion signal strength 
for $h_1$ production (red) and allowed regions derived from fits to the signal strengths given in Eq.
\ref{eq:mulims} (black)\cite{ATLAS-CONF-2015-044}
 with $\Lambda=2 ~TeV$.  Only the linear terms in the EFT expansion are included.}
\label{fig:muggf}
\end{figure}

The  $95\%$  confidence level limit from the $ggF$ signal strength is shown in Fig. \ref{fig:muggf}.
This fit 
demonstrates the cancellations between the contributions of the singlet model and the 
contributions of the EFT coefficients.   For $s_\theta=0$, the EFT operators do not contribute to $h_1$
decay, and so
there is  no limit on $c_{gg}$ (the lower band extending across all $c_{gg}$ values). For $s_\theta=1$, the
SM contributions vanish and the observed $h_1$ production rate is obtained by adjusting $c_{gg}$ (we 
have only plotted allowed values).   For small $c_{gg}$, we observe the interplay of the mixing and EFT
contributions, and larger values of $s_\theta$ are allowed than in the $c_{gg}=0$ limit.  In this plot, we
retain only the linear contributions in $c_{gg}$.  If the $c_{gg}^2$ terms become numerically relevant, then
the dimension-6 terms must be included in the EFT of Eq. \ref{eff}.

In Fig. \ref{fig:muggf},  we also fit the $h_1$ coupling strengths\cite{ATLAS-CONF-2015-044}
using the  $6$ parameter fit to the $gg$ initial state at $8~TeV$,
\begin{eqnarray}
\mu_F^{\gamma\gamma}&=&1.13^{+.24}_{-.21}\nonumber \\
\mu_F^{WW}&=&1.08^{.22}_{-.19} \nonumber \\
\mu_F^{ZZ}&=&1.29^{_.29}_{-.25} \nonumber \\
\mu_F^{bb}&=& .66 ^{+.37}_{-.28}\nonumber \\
\mu_F^{\tau\tau} &=& 1.07^{+.35}_{-.28}\, .
\label{eq:mulims}
\end{eqnarray}
These are labelled as ``h1 95\% CL Fits".  The results of the two fits are quite similar.  

\section{Constraints from $h_2$}
\label{sec:results}

We turn now to a joint examination of the measured properties of the $h_1$ as given in Eq. \ref{eq:mulims} and the experimental limits on
heavy resonances shown in Tabs. \ref{tab:reslims8} and \ref{tab:reslims13} for
heavy scalars with masses of $m_2=400,~600$ and $750~GeV$ decaying to SM particles using the results of Eq. \ref{h2dec}.
We calculate the signal rates at leading-order in QCD and normalize to the recommended values for the SM production rates from the LHC Higgs Cross Section Working Group\cite{Heinemeyer:2013tqa} given in Tab. \ref{tab:ressigs}.
\begin{table}
\begin{center}
\begin{tabular}{|c|c|c|c|}
\hline\hline
Channel & $m_2=400~\gev$ & $m_2=600~\gev$ & $m_2=750~\gev$\\
  \hline\hline 
 $WW$ &$0.362~\pb$\cite{Aad:2015agg}&$0.118~\pb$\cite{Aad:2015agg}
 & $0.0361~\pb$ \cite{Aad:2015agg} \\
 \hline
 $ZZ$ &$0.0648~\pb$\cite{Aad:2015kna} &
 $0.0218~\pb$\cite{Aad:2015kna}
 &$0.0118~\pb$\cite{Aad:2015kna} \\
 \hline
 $t {\overline t}$ 
 &*
 &$1.2~\pb$\cite{Aad:2015fna}
 &$0.71~\pb$\cite{Aad:2015fna}
\\
 \hline 
 $Z\gamma $ &$0.00720~\pb$\cite{Aad:2014fha}&$0.00296~\pb$\cite{Aad:2014fha}&$0.00402~\pb$\cite{Aad:2014fha}  \\
 \hline
 $\tau^+\tau^-$ &$0.087~\pb$
 \cite{Aad:2014vgg}&$0.020~\pb$\cite{Aad:2014vgg}& $0.012~\pb$\cite{Aad:2014vgg}  \\
 \hline
 $jj$
 &* & $3.76~\pb$\cite{Khachatryan:2016ecr}
 &
$1.79~\pb$ \cite{Khachatryan:2016ecr} \\
 \hline
 $h_1h_1$ &
 $0.442~\pb$\cite{Khachatryan:2015yea}&
 $0.137~\pb$\cite{Khachatryan:2015yea}& $0.0498~\pb$ \cite{Khachatryan:2015yea}
 \\
 \hline
 $\gamma\gamma$ & $0.00215~\pb$\cite{Khachatryan:2015qba}&$0.000666~\pb$\cite{Aad:2014ioa}&$0.00129~\pb$\cite{Khachatryan:2015qba}\\
  \hline\hline
  \end{tabular}
 \caption{\label{tab:reslims8} $95~\%$ c.l. LHC limits on  $\sigma\cdot  BR$ for heavy resonances at $\sqrt{S_H}=8~TeV$. } \end{center}
 \end{table}

\begin{table}
\begin{center}
\begin{tabular}{|c|c|c|c|}
\hline\hline
Channel & $m_2=400~\gev$ & $m_2=600~\gev$ & $m_2=750~\gev$\\
  \hline\hline 
 $WW$ &$1.4$ pb\cite{ATLAS-CONF-2016-074}&$0.5$ pb\cite{ATLAS-CONF-2016-074} &$0.31~\pb$\cite{ATLAS-CONF-2016-074} \\
 \hline
 $ZZ$ &$0.210~\pb$\cite{CMS-PAS-HIG-16-033}&$0.083~\pb$\cite{ATLAS-CONF-2016-056}&
 $0.043~\pb$\cite{ATLAS-CONF-2016-056}\\
 \hline
  $Z\gamma $ &$0.041~\pb$\cite{ATLAS-CONF-2016-044}&$.013~\pb$\cite{ATLAS-CONF-2016-044}&
 $0.010~\pb$\cite{ATLAS-CONF-2016-044}\\
 \hline
 $\tau^+\tau^-$ &$0.27~\pb$\cite{ATLAS-CONF-2016-085} &$0.053~\pb$\cite{ATLAS-CONF-2016-085} &
 $0.030~\pb$\cite{ATLAS-CONF-2016-085} \\
 \hline
 $jj$ &$*$ &$21.4~\pb$\cite{CMS-PAS-EXO-16-032}&$9.54~\pb$\cite{CMS-PAS-EXO-16-032}
 \\
  \hline
 $h_1h_1$ &$5.9~\pb$\cite{CMS-PAS-HIG-16-029}
 &$1.6~\pb$\cite{CMS-PAS-HIG-16-029}&
 $0.85~\pb$\cite{CMS-PAS-HIG-16-029}
 \\
 \hline
 $\gamma\gamma$ &$0.0018~\pb$\cite{ATLAS-CONF-2016-059}&
 $0.0015~\pb$\cite{ATLAS-CONF-2016-059}&
 $0.00068~\pb$\cite{ATLAS-CONF-2016-059}\\
 \hline
 $b{\overline{b}}$&$*$&$5.1~\pb$\cite{CMS-PAS-HIG-16-025}&$5.2~\pb$\cite{CMS-PAS-HIG-16-025}\\
  \hline\hline
  \end{tabular}
 \caption{\label{tab:reslims13} $95~\%$ C.L. LHC limits on  $\sigma\cdot  BR$ for heavy resonances at $\sqrt{S_H}=13~GeV$.} \end{center}
 \end{table}

 \begin{table}
\begin{center}
\begin{tabular}{|c|c|c|}
\hline\hline
& $8~\tev, \sigma(pp\rightarrow h_2)$ & $13~\tev, \sigma(pp\rightarrow h_2)$  \\
 \hline\hline 
$m_2=400~\gev$ & $3.01~\pb$&$9.52~\pb$\\
$m_2=600~\gev$&$0.52~\pb$&$2.01~\pb$\\
$m_2=750~\gev$ &$0.15~\pb$&$0.64~\pb$\\
  \hline\hline
  \end{tabular}
 \caption{\label{tab:ressigs} Theoretical  cross sections at NNLO+NNLL for heavy scalar resonances from the LHC Higgs
 Cross Section Working Group\cite{Heinemeyer:2013tqa}.}
  \end{center}
 \end{table}

\begin{figure}
\begin{centering}
\includegraphics[width=0.3\textwidth]{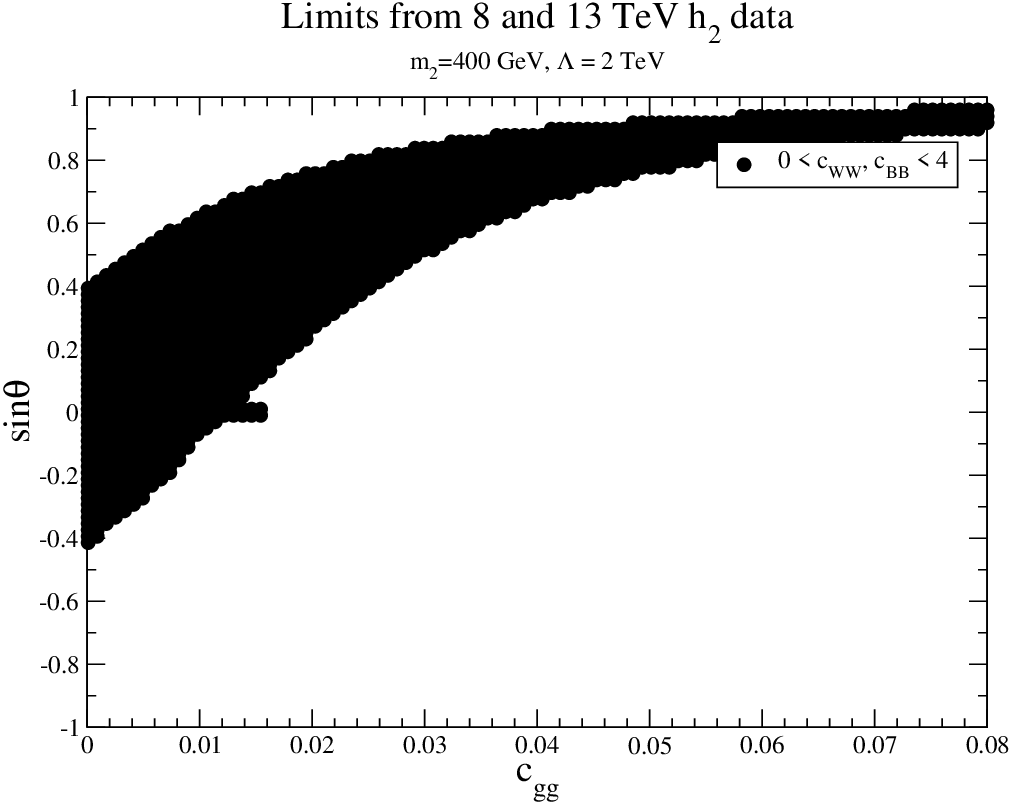}
\includegraphics[width=0.3\textwidth]{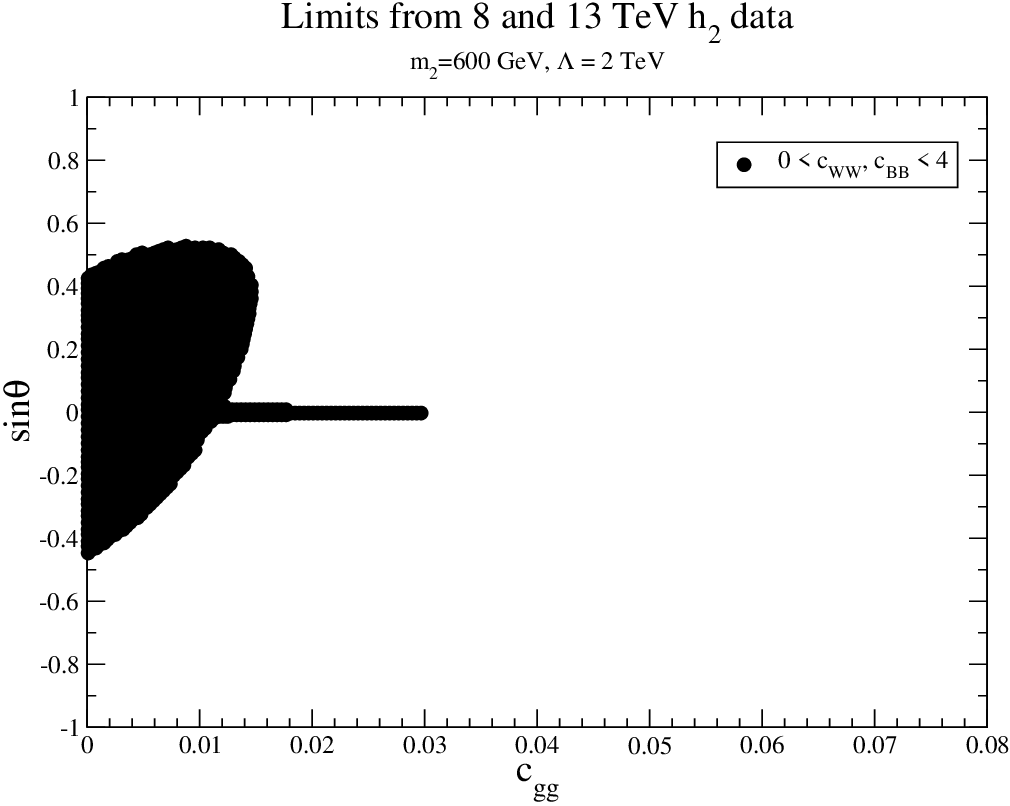}
\includegraphics[width=0.3\textwidth]{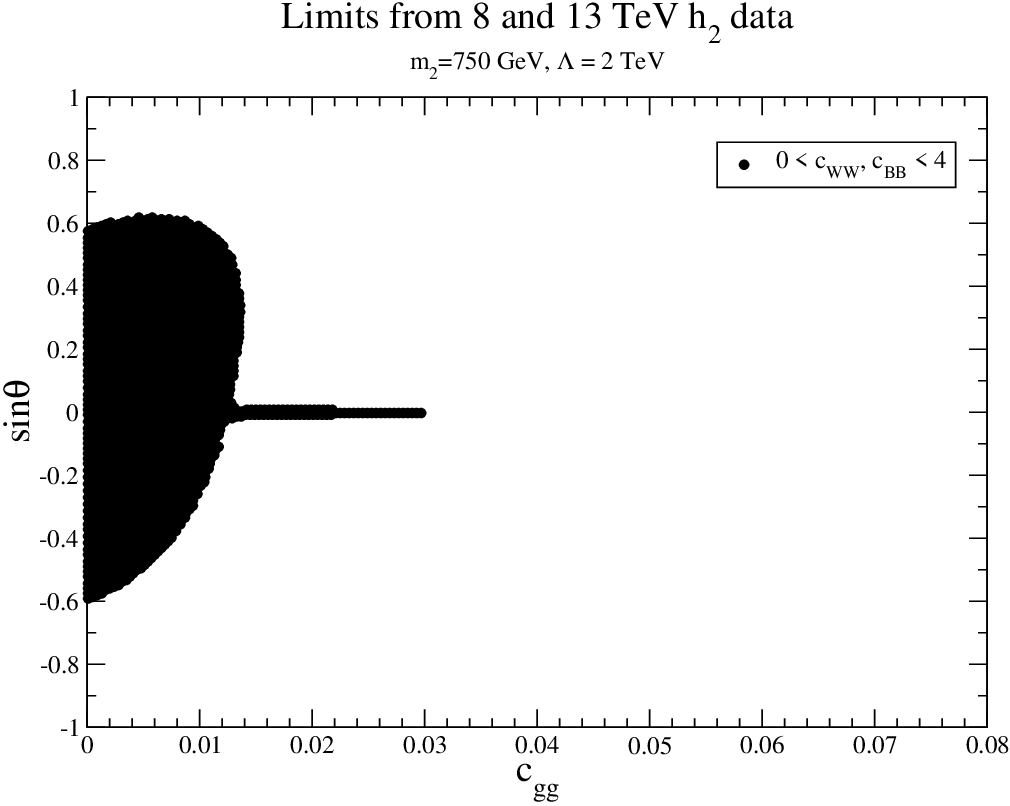}
\par\end{centering}
\caption{95\% confidence level allowed regions obtained by  varying $c_{gg}$, $c_{WW}$, $c_{BB}$,
$\cos\theta$, along with $b_1,b_3$ and $a_2$,  allowed by the $8~\tev$ 
and $13~\tev$ resonance searches of Tabs. \ref{tab:reslims8} and \ref{tab:reslims13}.
}
\label{fig:cggsth}
\end{figure}

Fig. \ref{fig:cggsth}  shows the regions excluded from  the 
the restrictions  from resonance searches
 at $8~\tev$ and $13~\tev$.  For $\sin\theta=0$, there is now an upper limit to $c_{gg}$ that arises from the
 dijet searches.    The region at $\sin\theta=1$, present in the $h_1$ fits, largely vanishes at $m_2=600$ and $750~\gev$, and
 is greatly reduced at $m_2=400~\gev$.     
 The excluded region shows little sensitivity to the parameter of the scalar potential. 
  The counting of
small parameters is different for the $h_2$ decays, than in the $h_1$ case. 
If we treat both $s_\theta$ and $c_i$ as small parameters, then
the $c_i^2$ contributions to $h_2$ decays are of the same order as the terms independent of the $c_i$.  Hence for the $h_2$ 
decays, we include the $c_i^2$ contributions.

\begin{figure}
\begin{centering}
\includegraphics[width=0.3\textwidth]{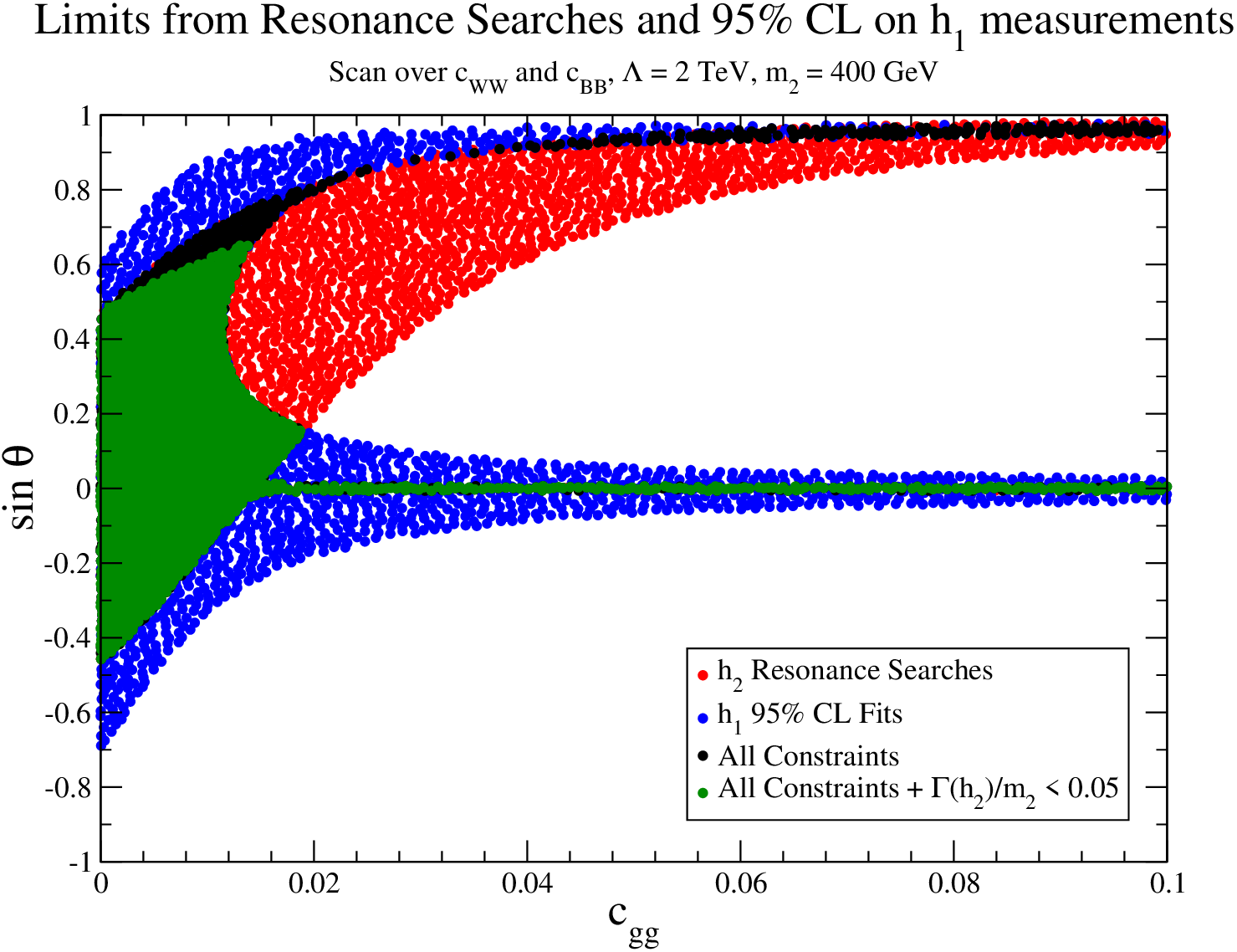}
\includegraphics[width=0.3\textwidth]{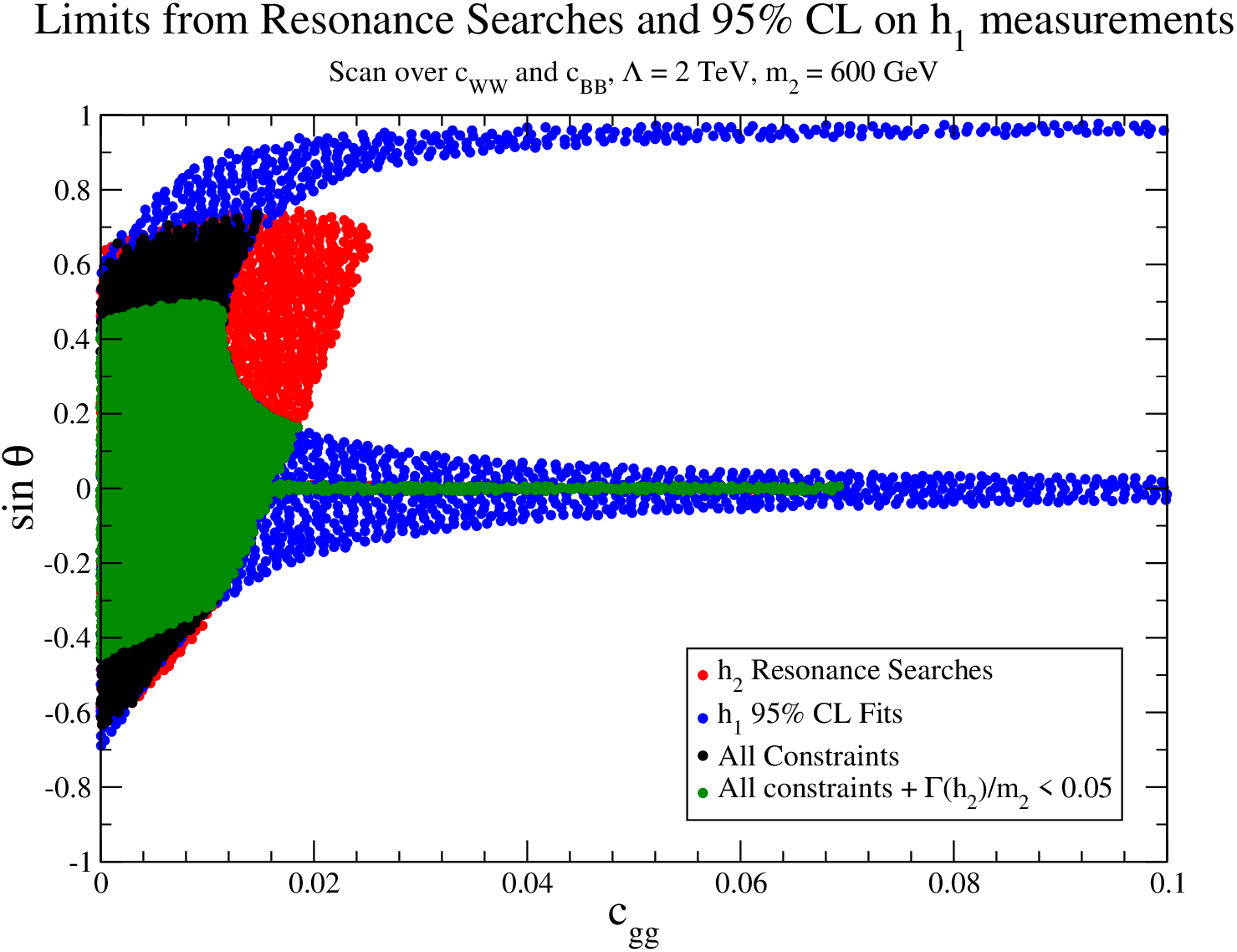}
\includegraphics[width=0.3\textwidth]{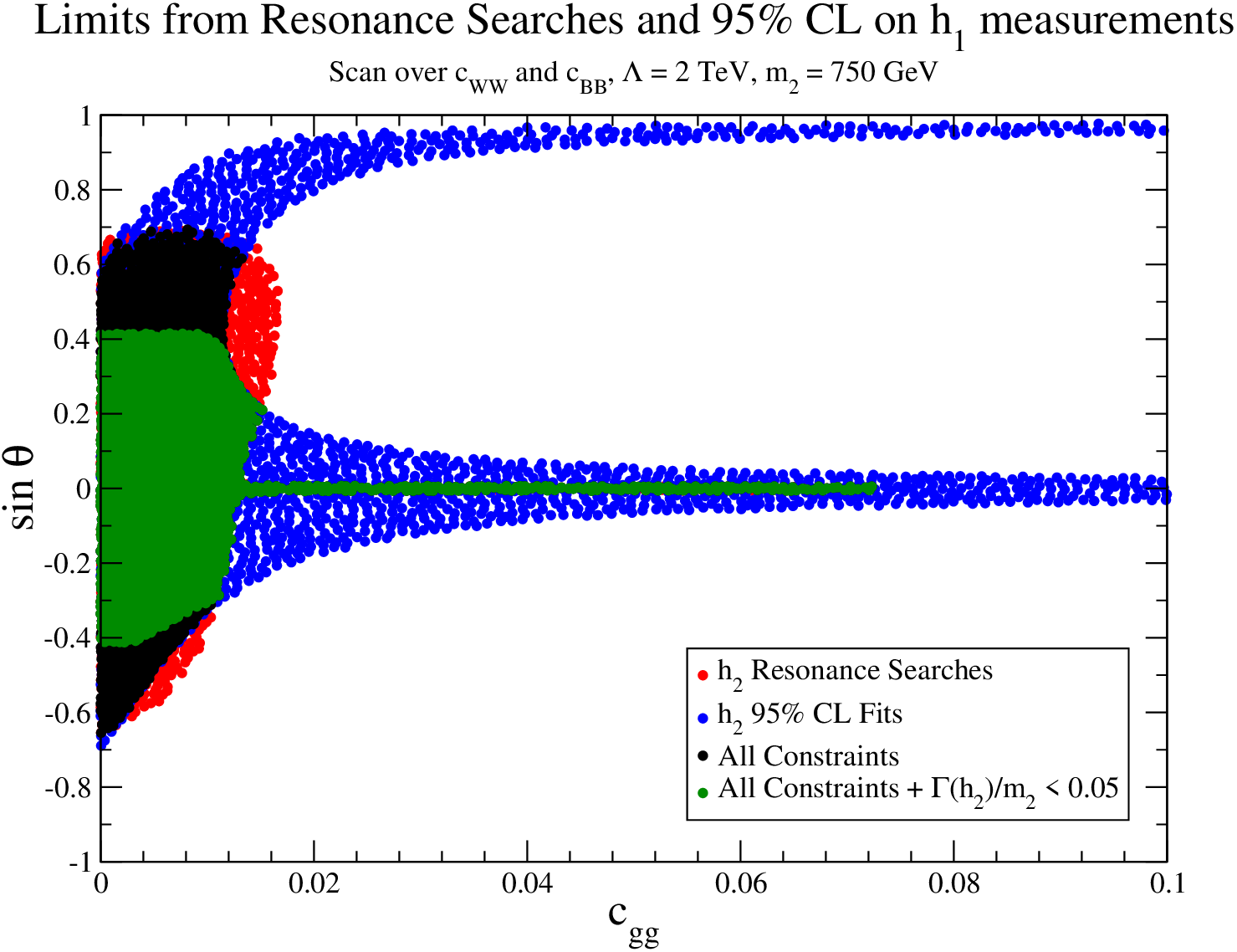}
\par\end{centering}
\caption{\label{fig:cggsth_all} Allowed regions combining $h_1$ and $h_2$ data and a narrow width $\Gamma(h_2)/m_2<0.05$ restriction.  The new physics scale is set $\Lambda=2$~TeV, and $c_{WW},c_{BB}$ are scanned over.}
\end{figure}

In Fig. \ref{fig:cggsth_all}, we plot the regions allowed by both $h_1$ coupling fits and resonance searches.  We see that
the large $c_{gg}$ regions that are allowed by the coupling constant fits are eliminated by the resonance search limits for $m_2=600$~GeV and $750$~GeV.   Considering all constraints, for $m_2=600$ and $750$~GeV we find $|\sin\theta|\lesssim 0.6$.  For $m_2=400$~GeV, the resonance searches are less restrictive for positive $\sin\theta$ and the limit is $\sin\theta\gtrsim -0.4$.  For all masses these limits are much weaker than $|\sin\theta|\leq 0.35$~\cite{Aad:2015pla} in the renormalizable model without the EFT operators in Eq.~\ref{eff}.

Finally, requiring a narrow width $\Gamma(h_2)/m_2<5\%$, where $\Gamma(h_2)$ is the total $h_2$ width, further constrains the allowed regions of $\sin\theta$.  For $m_2=600$ and $750$ GeV the limit is $|\sin\theta|\lesssim0.4$.  For $m_2=400$~GeV, the effect of the the narrow width restriction is to eliminate the large $\sin\theta\sim1$ region.  The remaining parameter region is $-0.4\lesssim \sin\theta\lesssim 0.7$.

\section{Conclusions}
We examined the effects on Higgs physics of  a gauge singlet scalar which mixes with the SM-like $125~GeV$ Higgs boson when the theory is augmented by EFT operators coupling the singlet scalar to SM gauge bosons.  
The new feature of our analysis is a  study of the properties of both the $125~GeV$ and
heavy scalar resonance, and the demonstration that significant cancellations are possible
between effects in the two sectors.  We fit our model parameters to the 7 and 8 TeV combined ATLAS and CMS precision Higgs measurements~\cite{ATLAS-CONF-2015-044} and applied constraints from scalar resonance searches at the 8 and 13 TeV LHC. 

We find that the inclusion of the operators greatly changes the allowed values of the scalar mixing angle.  In the renormalizable model, the strongest bound from Higgs precision is $|\sin\theta|\leq 0.35$~\cite{Aad:2015pla}.  Including the EFT operators between the singlet scalar and SM gauge bosons, we find Higgs precision measurements and scalar resonance searches give $\sin\theta\gtrsim -0.4$ for a heavy scalar mass of 400 GeV and $|\sin\theta|\lesssim0.6$ for masses of 600 and 750 GeV.  If the additional requirement of a narrow width $\Gamma(h_2)/m_2<0.05$ is included, the limits are $-0.4\lesssim \sin\theta\lesssim 0.7$ for a heavy scalar mass of 400 GeV and $|\sin\theta|\lesssim0.4$ for masses of 600 and 750 GeV.  In all cases, these restrictions are less than those in the renormalizable theory.

\section*{Acknowledgements}
This work is supported by the U.S. Department of Energy under grant
 DE-SC0012704.
We thank Chien-Yi Chen for many valuable discussions about the singlet
model. Digital data related to our results can be found at
$quark.phy.bnl.gov\backslash Digital\_Data\_Archive\backslash dawson\backslash singlet\_16$.

\newpage
\bibliographystyle{hunsrt}
\bibliography{hh750}

\begin{thebibliography}{10}

\bibitem{Espinosa:2011ax}
Jose~R. Espinosa, Thomas Konstandin, and Francesco Riva.
\newblock {Strong Electroweak Phase Transitions in the Standard Model with a
  Singlet}.
\newblock {\em Nucl.Phys.}, B854:592--630, 2012, 1107.5441.

\bibitem{Profumo:2007wc}
Stefano Profumo, Michael~J. Ramsey-Musolf, and Gabe Shaughnessy.
\newblock {Singlet Higgs phenomenology and the electroweak phase transition}.
\newblock {\em JHEP}, 0708:010, 2007, 0705.2425.

\bibitem{Profumo:2014opa}
Stefano Profumo, Michael~J. Ramsey-Musolf, Carroll~L. Wainwright, and Peter
  Winslow.
\newblock {Singlet-catalyzed electroweak phase transitions and precision Higgs
  boson studies}.
\newblock {\em Phys. Rev.}, D91(3):035018, 2015, 1407.5342.

\bibitem{Curtin:2014jma}
David Curtin, Patrick Meade, and Chiu-Tien Yu.
\newblock {Testing Electroweak Baryogenesis with Future Colliders}.
\newblock 2014, 1409.0005.

\bibitem{Perelstein:2016cxy}
Maxim Perelstein and Yu-Dai Tsai.
\newblock {750 GeV Di-photon Excess and Strongly First-Order Electroweak Phase
  Transition}.
\newblock 2016, 1603.04488.

\bibitem{O'Connell:2006wi}
Donal O'Connell, Michael~J. Ramsey-Musolf, and Mark~B. Wise.
\newblock {Minimal Extension of the Standard Model Scalar Sector}.
\newblock {\em Phys.Rev.}, D75:037701, 2007, hep-ph/0611014.

\bibitem{Barger:2008jx}
Vernon Barger, Paul Langacker, Mathew McCaskey, Michael Ramsey-Musolf, and Gabe
  Shaughnessy.
\newblock {Complex Singlet Extension of the Standard Model}.
\newblock {\em Phys.Rev.}, D79:015018, 2009, 0811.0393.

\bibitem{Pruna:2013bma}
Giovanni~Marco Pruna and Tania Robens.
\newblock {The Higgs Singlet extension parameter space in the light of the LHC
  discovery}.
\newblock {\em Phys.Rev.}, D88:115012, 2013, 1303.1150.

\bibitem{Chen:2014ask}
Chien-Yi Chen, S.~Dawson, and I.~M. Lewis.
\newblock {Exploring resonant di-Higgs boson production in the Higgs singlet
  model}.
\newblock {\em Phys. Rev.}, D91(3):035015, 2015, 1410.5488.

\bibitem{Robens:2016xkb}
Tania Robens and Tim Stefaniak.
\newblock {LHC Benchmark Scenarios for the Real Higgs Singlet Extension of the
  Standard Model}.
\newblock 2016, 1601.07880.

\bibitem{Dawson:2015haa}
S.~Dawson and I.~M. Lewis.
\newblock {NLO corrections to double Higgs boson production in the Higgs
  singlet model}.
\newblock {\em Phys. Rev.}, D92(9):094023, 2015, 1508.05397.

\bibitem{Costa:2015llh}
Raul Costa, Margarete Mühlleitner, Marco O.~P. Sampaio, and Rui Santos.
\newblock {Singlet Extensions of the Standard Model at LHC Run 2: Benchmarks
  and Comparison with the NMSSM}.
\newblock 2015, 1512.05355.

\bibitem{Ahriche:2007jp}
Amine Ahriche.
\newblock {What is the criterion for a strong first order electroweak phase
  transition in singlet models?}
\newblock {\em Phys.Rev.}, D75:083522, 2007, hep-ph/0701192.

\bibitem{Aad:2015pla}
Georges Aad et~al.
\newblock {Constraints on new phenomena via Higgs boson couplings and invisible
  decays with the ATLAS detector}.
\newblock {\em JHEP}, 11:206, 2015, 1509.00672.

\bibitem{ATLAS-CONF-2015-044}
{Measurements of the Higgs boson production and decay rates and constraints on
  its couplings from a combined ATLAS and CMS analysis of the LHC pp collision
  data at $\sqrt{s}$ = 7 and 8 TeV}.
\newblock Technical Report ATLAS-CONF-2015-044, CERN, Geneva, Sep 2015.

\bibitem{Dawson:2009yx}
Sally Dawson and Wenbin Yan.
\newblock {Hiding the Higgs Boson with Multiple Scalars}.
\newblock {\em Phys.Rev.}, D79:095002, 2009, 0904.2005.

\bibitem{Bauer:2016hcu}
Martin Bauer, Anja Butter, Juan Gonzalez-Fraile, Tilman Plehn, and Michael
  Rauch.
\newblock {Learning from the New Higgs-like Scalar before It Vanishes}.
\newblock 2016, 1607.04562.

\bibitem{Cheung:2016trn}
Kingman Cheung, P.~Ko, Jae~Sik Lee, Jubin Park, and Po-Yan Tseng.
\newblock {Double Higgcision: 125 GeV Higgs boson and a potential diphoton
  Resonance}.
\newblock 2016, 1608.00382.

\bibitem{Gunion:1989we}
John~F. Gunion, Howard~E. Haber, Gordon~L. Kane, and Sally Dawson.
\newblock {The Higgs Hunter's Guide}.
\newblock {\em Front. Phys.}, 80:1--448, 2000.

\bibitem{Djouadi:2005gi}
Abdelhak Djouadi.
\newblock {The Anatomy of electro-weak symmetry breaking. I: The Higgs boson in
  the standard model}.
\newblock {\em Phys. Rept.}, 457:1--216, 2008, hep-ph/0503172.

\bibitem{Contino:2014aaa}
Roberto Contino, Margherita Ghezzi, Christophe Grojean, Margarete Mühlleitner,
  and Michael Spira.
\newblock {eHDECAY: an Implementation of the Higgs Effective Lagrangian into
  HDECAY}.
\newblock {\em Comput. Phys. Commun.}, 185:3412--3423, 2014, 1403.3381.

\bibitem{Heinemeyer:2013tqa}
{LHC Higgs Cross Section Working Group}, S.~Heinemeyer, C.~Mariotti,
  G.~Passarino, and R.~Tanaka~(Eds.).
\newblock {Handbook of LHC Higgs Cross Sections: 3. Higgs Properties}.
\newblock {\em CERN-2013-004}, CERN, Geneva, 2013, 1307.1347.

\bibitem{Aad:2015agg}
Georges Aad et~al.
\newblock {Search for a high-mass Higgs boson decaying to a $W$ boson pair in
  $pp$ collisions at $\sqrt{s} = 8$ TeV with the ATLAS detector}.
\newblock {\em JHEP}, 01:032, 2016, 1509.00389.

\bibitem{Aad:2015kna}
Georges Aad et~al.
\newblock {Search for an additional, heavy Higgs boson in the $H\rightarrow ZZ$
  decay channel at $\sqrt{s} = 8\;\text{ TeV }$ in $pp$ collision data with the
  ATLAS detector}.
\newblock {\em Eur. Phys. J.}, C76(1):45, 2016, 1507.05930.

\bibitem{Aad:2015fna}
Georges Aad et~al.
\newblock {A search for $ t\overline{t} $ resonances using lepton-plus-jets
  events in proton-proton collisions at $ \sqrt{s}=8 $ TeV with the ATLAS
  detector}.
\newblock {\em JHEP}, 08:148, 2015, 1505.07018.

\bibitem{Aad:2014fha}
Georges Aad et~al.
\newblock {Search for new resonances in $W\gamma$ and $Z\gamma$ final states in
  $pp$ collisions at $\sqrt s=8$ TeV with the ATLAS detector}.
\newblock {\em Phys. Lett.}, B738:428--447, 2014, 1407.8150.

\bibitem{Aad:2014vgg}
Georges Aad et~al.
\newblock {Search for neutral Higgs bosons of the minimal supersymmetric
  standard model in pp collisions at $\sqrt{s}$ = 8 TeV with the ATLAS
  detector}.
\newblock {\em JHEP}, 11:056, 2014, 1409.6064.

\bibitem{Khachatryan:2016ecr}
Vardan Khachatryan et~al.
\newblock {Search for narrow resonances in dijet final states at sqrt(s)=8 TeV
  with the novel CMS technique of data scouting}.
\newblock 2016, 1604.08907.

\bibitem{Khachatryan:2015yea}
Vardan Khachatryan et~al.
\newblock {Search for resonant pair production of Higgs bosons decaying to two
  bottom quark?antiquark pairs in proton?proton collisions at 8 TeV}.
\newblock {\em Phys. Lett.}, B749:560--582, 2015, 1503.04114.

\bibitem{Khachatryan:2015qba}
Vardan Khachatryan et~al.
\newblock {Search for diphoton resonances in the mass range from 150 to 850 GeV
  in pp collisions at $\sqrt{s} =$ 8 TeV}.
\newblock {\em Phys. Lett.}, B750:494--519, 2015, 1506.02301.

\bibitem{Aad:2014ioa}
Georges Aad et~al.
\newblock {Search for Scalar Diphoton Resonances in the Mass Range $65-600$ GeV
  with the ATLAS Detector in $pp$ Collision Data at $\sqrt{s}$ = 8 $TeV$}.
\newblock {\em Phys. Rev. Lett.}, 113(17):171801, 2014, 1407.6583.

\bibitem{ATLAS-CONF-2016-074}
{Search for a high-mass Higgs boson decaying to a pair of $W$ bosons in $pp$
  collisions at $\sqrt{s}$=13 TeV with the ATLAS detector}.
\newblock Technical Report ATLAS-CONF-2016-074, CERN, Geneva, Aug 2016.

\bibitem{CMS-PAS-HIG-16-033}
{Measurements of properties of the Higgs boson and search for an additional
  resonance in the four-lepton final state at sqrt(s) = 13 TeV}.
\newblock Technical Report CMS-PAS-HIG-16-033, CERN, Geneva, 2016.

\bibitem{ATLAS-CONF-2016-056}
{Search for new phenomena in the $Z(\rightarrow\ell\ell) +
  E_{\mathrm{T}}^{\mathrm{miss}}$ final state at $\sqrt{s}$ = 13 TeV with the
  ATLAS detector}.
\newblock Technical Report ATLAS-CONF-2016-056, CERN, Geneva, Aug 2016.

\bibitem{ATLAS-CONF-2016-044}
{Search for new resonances decaying to a $Z$ boson and a photon in 13.3
  fb$^{-1}$ of $pp$ collisions at $\sqrt{s}=13$~TeV with the ATLAS detector}.
\newblock Technical Report ATLAS-CONF-2016-044, CERN, Geneva, Aug 2016.

\bibitem{ATLAS-CONF-2016-085}
{Search for Minimal Supersymmetric Standard Model Higgs Bosons $H/A$ in the
  $\tau\tau$ final state in up to 13.3 fb$^{?1}$ of pp collisions at
  $\sqrt{s}$= 13 TeV with the ATLAS Detector}.
\newblock Technical Report ATLAS-CONF-2016-085, CERN, Geneva, Aug 2016.

\bibitem{CMS-PAS-EXO-16-032}
{Searches for narrow resonances decaying to dijets in proton-proton collisions
  at 13 TeV using 12.9 inverse femtobarns.}
\newblock Technical Report CMS-PAS-EXO-16-032, CERN, Geneva, 2016.

\bibitem{CMS-PAS-HIG-16-029}
{Search for resonant Higgs boson pair production in the
  $\mathrm{b\overline{b}}\tau^+\tau^-$ final state using 2016 data}.
\newblock Technical Report CMS-PAS-HIG-16-029, CERN, Geneva, 2016.

\bibitem{ATLAS-CONF-2016-059}
{Search for scalar diphoton resonances with 15.4~fb$^{-1}$ of data collected at
  $\sqrt{s}$=13 TeV in 2015 and 2016 with the ATLAS detector}.
\newblock Technical Report ATLAS-CONF-2016-059, CERN, Geneva, Aug 2016.

\bibitem{CMS-PAS-HIG-16-025}
{Search for a narrow heavy decaying to bottom quark pairs in the 13 TeV data
  sample}.
\newblock Technical Report CMS-PAS-HIG-16-025, CERN, Geneva, 2016.

\end{thebibliography}

\end{document}